\documentclass{iopart}
\usepackage{graphicx}  
\usepackage{latexsym}
\usepackage{amssymb}
\usepackage{cite}

\jl{6}        % submitted to journal: Classical and Quantum Gravity
\eqnobysec    % section equation numbering

\def\beq{\begin{equation}}
\def\eeq{\end{equation}}
\def\rmd{{\rm d}}

\def\A{a}
\begin{document}

\title[Effect of radiation flux on test particle motion in the Vaidya spacetime]
{Effect of radiation flux on test particle motion in the Vaidya spacetime}

\author{
Donato Bini${}^*{}^\S{}^\ddag$, Andrea Geralico${}^\S$, 
Robert T. Jantzen${}^\P{}^\S$ and
Old\v{r}ich Semer\'ak${}^\mathdollar$
}
\address{
  ${}^*$\
Istituto per le Applicazioni del Calcolo ``M. Picone'', CNR I--00161 Rome, Italy
}
\address{
  ${}^\S$\
  International Center for Relativistic Astrophysics -- I.C.R.A.,
  University of Rome ``La Sapienza'', I--00185 Rome, Italy
}
\address{
  ${}^\ddag$\
  INFN sezione di Firenze,
  Sesto Fiorentino (FI), Italy
}
\address{
  ${}^\P$\
Department of Mathematical Sciences, Villanova University, Villanova, PA 19085, USA
}
\address{
  ${}^\mathdollar$\
Institute of Theoretical Physics, Faculty of Mathematics and Physics, 
Charles University in Prague, Czech Republic
}  

\begin{abstract}
Motion of massive test particles in the nonvacuum spherically symmetric radiating Vaidya spacetime is investigated, allowing for physical interaction of the particles with the radiation field in terms of which the source energy-momentum tensor is interpreted. This ``Poynting-Robertson-like effect" is modeled by the usual effective term describing a Thomson-type radiation drag force. The equations of motion are studied for simple types of motion including free motion (without interaction), purely radial and purely azimuthal (circular) motion, and for the particular case of ``static" equilibrium; appropriate solutions are given where possible. The results---mainly those on the possible existence of equilibrium positions---are compared with their counterparts obtained previously for a test spherically symmetric radiation field in a vacuum Schwarzschild background.
\end{abstract}

\pacno{04.20.Cv}

\section{Introduction}

Test particle motion in realistic gravitational fields is of obvious astrophysical importance and at the same time it provides reliable evidence of the properties of those gravitational fields.
However, in many actual astrophysical systems the particles are not moving freely but are influenced by ambient matter, electromagnetic fields and radiation. In typical situations, these ``physical" effects are probably even more important than fine details of the spacetime geometry alone. The most remarkable conditions, from the point of view of general relativity as well as astrophysics, appear near very compact objects where both the pure gravitational and other ``physical" effects typically become extraordinarily strong.

In the present paper we focus on the motion of test particles in a spherically symmetric gravitational field, under the action of a Thomson-type interaction with radiation emitted or accreted by a compact center. This kind of problem was first investigated by Poynting \cite{Poynting-03} using Newtonian gravity and then in the framework of linearized general relativity by Robertson \cite{Robertson-37}. It involves competition between gravity and radiation drag, which may lead to interesting types of motion which do not occur in strictly vacuum circumstances. In particular, there arises the question of whether equilibrium behavior like circular orbit motion or even ``staying at rest" are possible in some cases. Theoretical aspects of the Poynting-Robertson effect as well as its astrophysical relevance in specific situations have been studied by many authors since the original pioneering work. Recently we considered this same effect in the relativistic setting by studying
test particles orbiting in the equatorial plane of a Schwarzschild or Kerr black hole, assuming that the source of radiation is located symmetrically not far from the horizon (in the case of outgoing flux). We first chose the radiation field to be directed purely radially with respect to the zero-angular-momentum observers so that it too had zero angular momentum \cite{BiniJS-09}, but then also considered a more general case of radiation having some (arbitrary) angular momentum \cite{BiniGJSS-11} (see these papers for a more thorough overview of the references).
While ingoing radiation might at first seem rather unmotivated, its consideration can give some rough idea about particle motion inside an accretion disk with strong radiation emanating from the disk, at least in the plane of the disk.

Here we intend to compare the limiting simpler case of the Poynting-Robertson effect due to a purely radial (zero-angular-momentum) test radiation flux in a Schwarzschild background with the treatment using a self-consistent radiation flux in the exact, Vaidya spherically symmetric spacetime whose source includes a null dust  \cite{Vaidya-43,Vaidya-51,Vaidya-53}. 
Similarly one could think of comparing the general nonzero angular momentum test flux case in the Kerr spacetime background with the rotating Kerr-Vaidya spacetime \cite{vaidya-patel}, but the energy-momentum tensor of this latter exact solution was shown not to be interpretable in terms of a  null dust alone as in the nonrotating case \cite{Carmeli-Kaye}, so we will not address that issue here.
Since the simpler Vaidya spacetime contains an arbitrary function $M$ which describes a time-dependent mass for the central object, there is a freedom to choose that function in many ways, but we limit our attention to a relatively simple evolution of this function in our applications. In particular when an appropriate derivative of this function $M$ is constant, one models a phase of evolution of the central object in which the central mass changes at a constant rate. In this nonstatic situation, it no longer makes sense to seek radial equilibrium orbits as in the corresponding static test flux case, but one can look for some kind of adiabatically changing equivalent orbits.

After summarizing some basic properties of the Vaidya spacetime, we write out and reduce the equations describing timelike motion (with an effective force term describing the particle-radiation interaction) in Schwarzschild-like spherical coordinates with retarded/advanced time. 
Next we examine the appropriate limit of the Vaidya spacetime which leads to the scenario of a test flux in the Schwarzschild spacetime, in particular to see how the Schwarzschild radial equilibrium orbits fit into the more general situation.
Then several numerical examples are shown illustrating typical types of motion.

%%%%%%%%%%%%%%%%%%%%%%%%%%%%%%%%%%%%%%%%%%%%%%%%%%%%%%%%%%%%%%%%%%%%%%%

\section{Vaidya spacetime}
\label{Vaidya-summary}

The gravitational field associated with a spherically symmetric body of variable mass $M$ is described by Vaidya's nonvacuum solution of the Einstein equations \cite{Vaidya-43,Vaidya-51,Vaidya-53}, representing an algebraically special Petrov type D spacetime like the Schwarzschild metric it generalizes.
The corresponding line element written in Schwarzschild-like coordinates is given by 
\beq
\label{metrictr}
\fl\qquad
\rmd s^2=-\left(\frac{M_{,t}}{M_{,r}}\right)^2 \frac{{\rm d}t^2}{1-{2M}/{r}}
+\frac{{\rm d}r^2}{1-{2M}/{r}}+r^2(\rmd \theta^2+\sin^2\theta \rmd \phi^2)\,,
\eeq
with $M=M(t,r)$, and $M_{,t}$ and $M_{,r}$ denote partial derivatives with respect to $t$ and $r$ respectively.
A more useful form is obtained by introducing Eddington-Finkelstein-like coordinates through the transformation
to the null coordinate $u$ by
\beq
\rmd u=\mp\frac{r}{r-2M}\,\frac{{\rm d}M}{M_{,r}}\,,
\eeq
so that the line element becomes
\beq
\label{metricur}
\rmd s^2=-N^2\rmd u^2
\mp 2\rmd u \rmd r
+r^2(\rmd \theta^2+\sin^2\theta \rmd \phi^2)\,, 
\eeq
where 
$
N=\sqrt{1-{2M(u)}/{r}}
$.
Hereafter we will use the Eddington-Finkelstein-like form (\ref{metricur}). Its inverse has non-zero components
\beq
  g^{ur}=g_{ur}, \;\;\;
  g^{rr}=-g_{uu}, \;\;\;
  g^{\theta\theta}=(g_{\theta\theta})^{-1}, \;\;\;
  g^{\phi\phi}=(g_{\phi\phi})^{-1},
\eeq
so the raising or lowering of indices involves the simple relations $X_r=g_{ur}X^u=\mp X^u$.
In the case of non-positive/non-negative $M_{,u}\equiv {\rm d}M /{\rm d}u\,$, i.e., for a source loosing/gaining energy, $u$ represents retarded/advanced time and one has $g_{ur}=\mp 1$, respectively.\footnote{
In the entire article, upper signs correspond to $M_{,u}\leq 0$ (outgoing radiation) while lower signs to $M_{,u}\geq 0$ (ingoing radiation), so that $\pm M_{,u}\leq 0$. The advanced null coordinate is conventionally denoted by $v$, but we keep $u$ in both cases (distinguishing between them by signs).
}
We will allow for both possibilities---in fact we leave the $M(u)$ evolution completely general, with several important specific cases treated as examples.

The metric (\ref{metricur}) has a curvature singularity at $r=0$ and a coordinate singularity at $r=2M$ which corresponds to an apparent (but not event) horizon. Actually, on the $r=2M$ hypersurface the metric reduces to
\[{\rm d}s^2|_{r=2M}=
  \mp 4M_{,u}{\rm d}u^2+4M^2(\rmd \theta^2+\sin^2\theta \rmd \phi^2)\]
which is spacelike in both cases, i.e., with $\mp M_{,u}>0$.
In the $M={\rm const}$ limit the metric (\ref{metricur}) reduces to the Schwarzschild metric written in terms of the retarded/advanced time
\[u=t\mp\left[r+2M\ln\left(\frac{r}{2M}-1\right)\right]\,\]
for which $\partial_u=\partial_t$ and $\partial_\phi$ are both Killing vector fields.

Global properties of the Vaidya spacetime (its conformal structure) are summarized in section 9.5 of \cite{GriffithsP-09} together with the most relevant references. See also \cite{Poisson-04} where different definitions of mass (section 4.3.5) and the distinction between apparent and event horizons (section 5.1.8) are illustrated using the Vaidya spacetime as an example.

Evaluating the Einstein equations $G_{\mu\nu}=8\pi T_{\mu\nu}$ for this metric, one finds that the energy-momentum tensor has only one non-zero component in these coordinates
\beq  \label{Tmunu}
  T_{uu}=T^{rr}=\mp\frac{M_{,u}}{4\pi r^2} > 0 \,.
\eeq
Such a $T^{\mu\nu}$ can be interpreted as a null dust representing ``pure radiation," namely $T^{\mu\nu}=\Phi^2 k^\mu k^\nu$, where $k^\mu$ is a purely radial outgoing/ingoing null vector and $\Phi^2$ depends on the normalization chosen for $k^\mu$. For example, if one takes $k^\mu=\pm 2^{-1/2}\delta^\mu{}_r\,$, then
\beq
  T^{rr}=\frac{\Phi^2}{2} \quad\ \Longrightarrow \quad\
  \Phi^2=\mp\frac{M_{,u}}{2\pi r^2} \,.
\eeq
This choice for $k$ means that
$k_\alpha=g_{\alpha\mu}k^\mu=g_{\alpha r}k^r=\mp k^r\delta^u{}_\alpha=-2^{-1/2}\delta^u{}_\alpha$;
in particular, the energy of the radiation particles is proportional to $-k_u=\pm k^r=1/\sqrt{2}=E$. 
In the corresponding Schwarzschild spacetime with this value for $E$, $k$ agrees with the choice used in  \cite{BiniJS-09}. As in that case,
one checks easily that $k$ is tangent to a congruence of affinely parametrized null geodesics since
\[{k^\mu}_{;\nu}k^\nu
  ={\Gamma^\mu}_{\nu\lambda}k^\nu k^\lambda
  =\frac{1}{2}\,{\Gamma^\mu}_{rr}
  =0\,.
\]
This congruence has nonzero expansion ${k^\mu}_{;\mu}=\pm\sqrt{2}/r$ but zero vorticity.

The most natural test observers suitable for physical interpretation are those at rest in the spatial coordinate grid at 
$r={\rm const}$, $\theta={\rm const}$, $\phi={\rm const}$; their 4-velocity field is
\beq  \label{static-obs}
  \hat{u}\equiv e_{\hat u}=
  \frac{1}{N}\,\partial_u\,.
\eeq
A convenient spatial orthonormal triad tied to this observer congruence is
\beq
\label{framestatic}
  e_{\hat r}=N\left(\partial_r \mp
                    \frac{1}{N^2}\,\partial_u\right),\
  e_{\hat\theta}=\frac{1}{\sqrt{g_{\theta\theta}}}\,\partial_\theta\,,\
  e_{\hat\phi}=\frac{1}{\sqrt{g_{\phi\phi}}}\,\partial_\phi\,.
\eeq
This observer congruence is accelerated and has nonzero expansion and shear,
\begin{eqnarray}
\fl\qquad
  \lefteqn{
  a(\hat{u})^\mu\equiv\hat{u}^\mu{}_{;\nu}\hat{u}^\nu
    =\frac{MN^2\mp M_{,u}r}{r^2 N^3}\;e_{\hat r}^\mu \,,}\\
\fl\qquad
  \lefteqn{
  \theta({\hat u})^{\mu\nu}\equiv\hat{u}^{\mu;\nu}+a(\hat{u})^\mu\hat{u}^\nu
    =\hat\theta\,e_{\hat r}^\mu e_{\hat r}^\nu \,, \quad
  \hat\theta\equiv\hat{u}^\mu{}_{;\mu}
    =\frac{M_{,u}}{rN^3} \;,} \label{expansion}
\end{eqnarray}
but its vorticity vanishes,
\beq
  \omega(\hat{u})^{\mu\nu}\equiv\hat{u}^{[\mu;\nu]}+a(\hat{u})^{[\mu}\hat{u}^{\nu]}=0 \,,
\eeq
so it is hypersurface orthogonal. 
In fact these observers follow the $t$ coordinate lines orthogonal to the $t$ coordinate hypersurfaces in the original coordinate system, and in the Schwarzschild case $M_{,u} =0$ they are just the usual static observers.

\section{Test-particle motion and the Poynting-Robertson-like effect}
\label{motion-in-Vaidya}

When studying motion of massive test particles (rest mass $m\neq0$) in the Vaidya spacetime, one can either restrict attention to geodesics, or allow the particles to interact physically with the radiation in terms of which the energy-momentum tensor is interpreted.
A simple way to model this interaction is to 
assume that the force on the particle is proportional to the 4-momentum density of radiation observed in the particle's rest frame. 
 Denoting the particle's 4-velocity by $U^\alpha$ and acceleration by $a(U)^\alpha\equiv{\rm D}U^\alpha/{\rm d}\tau$, this yields the equation of motion
\beq  \label{motion}
  ma(U)^\alpha=-\sigma P(U)^\alpha{}_\mu T^\mu{}_\nu U^\nu
         \equiv{\cal F}_{\rm rad}(U)^\alpha\,,
\eeq
where $\tau$ and $m$ are particle's proper time and rest mass, $\sigma$ is the effective interaction cross section (its dimension is length squared) and $P(U)^\alpha{}_\mu\equiv \delta^\alpha{}_\mu+U^\alpha U_\mu$ is the projector to the particle's instantaneous rest space. Such a force formula is independent of the direction and frequency of radiation (and the interaction ``efficiency" $\sigma$ is also independent of the momentum-density magnitude) and can perhaps be adequate in situations when Thomson scattering is a dominant interaction mode.

Let us divide the equation (\ref{motion}) by $m$ (denoting $\tilde\sigma\equiv\sigma/m$) and write it out in the $(u,r,\theta,\phi)$ coordinates. Since the energy-momentum tensor has only one non-zero component $T_{uu}$ (\ref{Tmunu}), one has
\beq  \label{motion/m}
  \frac{{\rm d}U^\alpha}{{\rm d}\tau}=
    -{\Gamma^\alpha}_{\kappa\lambda}U^\kappa U^\lambda
    -\tilde\sigma(g^{\alpha u}+U^\alpha U^u)\,T_{uu}U^u \,.
\eeq
First, the latitudinal component
\beq
  \frac{{\rm d}U^\theta}{{\rm d}\tau}=
    -\frac{2}{r}\,U^r U^\theta+(U^\phi)^2\sin\theta\cos\theta
    -\tilde\sigma T_{uu}(U^u)^2 U^\theta
\eeq
confirms that the interaction leaves the motion planar: choosing $\theta=\pi/2$ and $U^\theta=0$ at some instant, one sees that at that instant ${\rm d}U^\theta/{\rm d}\tau=0$ too and so the particle remains in the (``equatorial") plane. Substituting $\theta=\pi/2$ and $U^\theta=0$ into the other components of (\ref{motion/m}), we have
\begin{eqnarray}
  \frac{{\rm d}U^u}{{\rm d}\tau}&=&
    \pm\frac{M}{r^2}\,(U^u)^2\mp r(U^\phi)^2-\tilde\sigma T_{uu}(U^u)^3 \,, \label{Uu-contra} \\
  \frac{{\rm d}U^r}{{\rm d}\tau}&=&
    -\left(\frac{M}{r}\,N^2\mp M_{,u}\right)\frac{(U^u)^2}{r}
    \mp\frac{2M}{r^2}\,U^u U^r +rN^2(U^\phi)^2\nonumber \\ 
    &&-\tilde\sigma(U^u U^r\mp 1)\,T_{uu}U^u \,,  \label{Ur-contra} \\
  \frac{{\rm d}U^\phi}{{\rm d}\tau}&=&
    -\frac{2}{r}\,U^r U^\phi
    -\tilde\sigma T_{uu}(U^u)^2 U^\phi \,.  \label{Uphi-contra}
\end{eqnarray}
The 4-velocity normalization condition $-1=g_{\mu\nu}U^\mu U^\nu$, explicitly
\beq
  -1 = -N^2(U^u)^2\mp 2U^u U^r+r^2(U^\phi)^2 \label{norm-contra} \,,
\eeq
enables the simplification of the longest equation (\ref{Ur-contra}) to
\beq  \label{Ur-contra-shorter}
\fl\qquad
  \frac{{\rm d}U^r}{{\rm d}\tau}=
    -\frac{M}{r^2}+(r-3M)(U^\phi)^2
    \pm\frac{M_{,u}}{r}\,(U^u)^2
    -\tilde\sigma(U^u U^r\mp 1)\,T_{uu}U^u \,.
\eeq
One could instead use the normalization condition to eliminate one of the three 4-velocity components from all of the equations
trying to make the latter a closed system, but unfortunately, even with a ``favorable" evolution of the mass $M=M(u)$, exact integration of such a system is almost never possible due to the a priori unknown dependence $r=r(\tau)$. In any case, the most important ``non-Schwarzschild" feature (besides the interaction terms scaled by $\tilde\sigma$) is the third term $\pm\frac{1}{r}\,M_{,u}(U^u)^2$ in the radial equation (\ref{Ur-contra-shorter}). It is clearly never positive, so it always increases the static radial pull $-M/r^2$. In order to balance this inward pull, the azimuthal velocity $U^\phi$ in the ``centrifugal" term $(r-3M)(U^\phi)^2$ has to be larger than in the Schwarzschild field. Finally, the last term of Eq.~(\ref{Ur-contra-shorter}) represents physical interaction of the particle with radiation; given that $\tilde\sigma T_{uu}U^u\geq 0$, it is seen that only in the case of outgoing radiation (upper sign) can this term oppose the gravitational attraction.

To interpret the test particle 4-velocity, one can express it using the obvious physical tetrad of
Eqs. (\ref{static-obs}), (\ref{framestatic})
adapted to the ``static" observer $\hat u$,
\beq
U=\gamma(U,\hat u) [\hat u + \nu(U, \hat u)^{\hat a}e_{\hat a}]\,,
\eeq
where $\gamma(U,\hat u)$ is the Lorentz factor
\beq
\gamma(U,\hat u)=({1-\delta_{\hat a \hat b}\,\nu(U, \hat u)^{\hat a}\nu(U, \hat u)^{\hat b} })^{-1/2}
  =  NU^u \pm \frac{1}{N}\,U^r
\eeq
and $e_{\hat a}$ are defined in Eq.~(\ref{framestatic}). 
The $4$-velocity tetrad components are related to the corresponding coordinate components by
\beq \label{eq:firstordervelocities}
\fl
\gamma(U,\hat u)\nu(U, \hat u)^{\hat r} = \frac{1}{N}\,U^r \,, \ 
\gamma(U,\hat u)\nu(U, \hat u)^{\hat \theta} = rU^\theta \,, \ 
\gamma(U,\hat u)\nu(U, \hat u)^{\hat \phi} = rU^\phi\sin\theta \,.
\eeq

For a particle moving in the equatorial plane $\theta=\pi/2$, so that $\nu(U,\hat u)^{\hat \theta}=0=U^\theta$, the equations of motion~(\ref{motion}) become
\begin{eqnarray}
\label{motion2}
\fl\quad
\frac{\rmd \nu^{\hat r}}{\rmd \tau}&=&
  -\frac{\gamma}{ rN^3}(1-(\nu^{\hat r}){}^2)\left[\frac{M}{r}\,N^2\mp M_{,u} (1\mp\nu^{\hat r}) \right] +\frac{\gamma N}{r}(\nu^{\hat \phi}){}^2
  \pm\frac{\tilde \sigma T_{uu}}{N^2}(1\mp\nu^{\hat r})^2\,,\nonumber \\
\fl\quad
\frac{\rmd \nu^{\hat \phi}}{\rmd \tau}&=&
  \frac{\gamma}{rN^3}\nu^{\hat r}\nu^{\hat \phi}
  \left[\frac{M}{r}\,N^2\mp M_{,u} (1\mp\nu^{\hat r})\right]-\frac{\gamma N}{r}\nu^{\hat r}\nu^{\hat \phi}-\frac{\tilde \sigma T_{uu}}{N^2}\nu^{\hat \phi}(1\mp\nu^{\hat r})\,,
\end{eqnarray}
with $\gamma(U, \hat u)\equiv \gamma=1/\sqrt{1-(\nu^{\hat r}){}^2-(\nu^{\hat \phi}){}^2}$.
To complete this system one must add the evolution equations for $u$, $r$ and $\phi$, i.e.,
\beq
\label{motionaltre3}
\frac{\rmd u}{\rmd \tau} = \frac{\gamma}{N}(1\mp\nu^{\hat r})\,,\quad
\frac{\rmd r}{\rmd \tau} = \gamma N \nu^{\hat r}\,,\quad 
\frac{\rmd \phi}{\rmd \tau} = \frac{\gamma \nu^{\hat \phi}}{r} \,.
\eeq

\section{Special types of motion}
\label{special-motions}

In general the above system of equations is only solvable numerically, but we will at least try to reduce it further for several particular, simple cases. First we will restrict to geodesic motion, then to purely radial ($\phi={\rm const}$) and to ``circular" ($r={\rm const}$) motion, and finally we will check whether equilibrium ($U^i=0$) between the gravitational pull and the radiation effect is possible.

\subsection{Free motion ($\tilde\sigma=0$)}
\label{geodesics}

First we check how the central mass change itself affects the free (geodesic) motion of test particles, not taking the interaction with radiation into account ($\tilde\sigma=0$); this problem was discussed in \cite{LindquistSM-65}. The most important simplification arises due to the azimuthal symmetry of the field: the specific angular momentum $\tilde{L}\equiv L/m\equiv U_\phi$ is a constant of the motion as in the Schwarzschild spacetime, so one has 
\beq
  U^\phi=g^{\phi\phi}U_\phi=\frac{\tilde{L}}{r^2}
\eeq
immediately and need not to solve the ``azimuthal" Eq.~(\ref{Uphi-contra}). Equations for the remaining two 4-velocity components read
\begin{eqnarray}
  \frac{{\rm d}U^u}{{\rm d}\tau}&=&
    \pm\frac{M}{r^2}\,(U^u)^2\mp\frac{\tilde{L}^2}{r^3} \,,\\
  \frac{{\rm d}U^r}{{\rm d}\tau}&=&
    -\frac{M}{r^2}+(r-3M)\,\frac{\tilde{L}^2}{r^4}
    \pm\frac{M_{,u}}{r}\,(U^u)^2\,
\end{eqnarray}
in contravariant form, while the covariant form reduces to
\begin{eqnarray}
  \frac{{\rm d}U_u}{{\rm d}\tau}&=&\frac{M_{,u}}{r}\,(U_r)^2
      \left(=\mp\frac{{\rm d}M}{{\rm d}\tau}\,\frac{U_r}{r}
            =\frac{{\rm d}M}{{\rm d}\tau}\,\frac{U^u}{r}\right), \label{dUu-geod} \\
  \frac{{\rm d}U_r}{{\rm d}\tau}&=&-\frac{M}{r^2}\,(U_r)^2+\frac{\tilde{L}^2}{r^3} \,.
\end{eqnarray}
Since $-U_u$ represents the particle's energy, Eq.~(\ref{dUu-geod}) says that a free particle gains/looses energy if $M$ decreases/increases (see \cite{DenisovaZ-00}). It must be stressed that this has nothing to do with the radiation (there is no interaction in the case of geodesics), the energy increases/decreases simply because the particle's gravitational binding by the center weakens/strengthens.
Formally the geodesic equations only differ from the Schwarzschild case by the $M_{,u}$ term in the radial equation, but one must always remember that $M$ itself depends on $u$ and thus also changes with $\tau$ everywhere.

\subsection{Purely radial motion ($\tilde\sigma\neq0$, $U^\phi=0$)}

If $U^\phi=0$ at some instant, one sees from Eq.~(\ref{Uphi-contra}) that ${{\rm d}U^\phi}/{{\rm d}\tau}=0$ there as well, so $U^\phi$ (and also $U_\phi$) remains zero along the whole world line. Equations for the time and radial components of motion are then
\begin{eqnarray}
  \frac{{\rm d}U^u}{{\rm d}\tau}&=&
    \pm\frac{M}{r^2}\,(U^u)^2-\tilde\sigma T_{uu}(U^u)^3 \,,\\
  \frac{{\rm d}U^r}{{\rm d}\tau}&=&
    -\frac{M}{r^2}\pm\frac{M_{,u}}{r}\,(U^u)^2-\tilde\sigma(U^u U^r\mp 1)\,T_{uu}U^u\,.
\end{eqnarray}
Substituting for $T_{uu}$ (\ref{Tmunu}) and writing it in terms of $M_{,u}U^u={\rm d}M/{\rm d}\tau$, the first equation acquires the form
\beq  \label{dUu-radial}
  \frac{{\rm d}U^u}{{\rm d}\tau}=
  \pm\frac{(U^u)^2}{r^2}\left(M+\frac{\tilde\sigma}{4\pi}\frac{{\rm d}M}{{\rm d}\tau}\right).
\eeq
Here all the quantities are treated as functions of $\tau$. In particular, the evolution of the radius $r(\tau)$ is of course not known a priori, and so the system is analytically unsolvable in general. One can, however, obtain a solution for particular choices of $M(\tau)$.
For example, for an exponential decay of the mass, $M(\tau)=M(0)\exp(-4\pi\tau/\tilde\sigma)$, the term in parentheses in (\ref{dUu-radial}) vanishes, which implies $U^u={\rm const}$. With $U^r$ replaced using the normalization condition, one finds
\beq
  \frac{{\rm d}r}{{\rm d}\tau}\equiv U^r
  =(2U^u)^{-1}\left[1-N^2(U^u)^2\right]\,,
\eeq
which can be solved numerically for $r(\tau)$. In particular, for the choice $U^u=1$, this is exactly solved by
\[r^2(\tau)=r_0{}^2+
            \frac{\tilde\sigma}{4\pi}\,M_0
                  \left[1-\exp\!\left(\!-\frac{4\pi\tau}{\tilde\sigma}\!\right)\right]\]
which represents the growth of $r$ from $r_0\equiv r(0)$ to an asymptotic value
$\sqrt{r_0{}^2+M_0\,\tilde\sigma/(4\pi)}\,$. However, it is clear that this is just one ad hoc, artificial case.

The system is analytically unsolvable even in the geodesic limit ($\tilde\sigma=0$), although it becomes quite compact in that case. The geodesic form of the radial equation is worth mentioning, in particular: after substituting $N^2\,{(U^u)^2}=1\mp 2U^u U^r$ from the normalization, it reads
\beq  \label{ddr,time-like}
  rN^2\,\frac{{\rm d}^2 r}{{\rm d}\tau^2}
  =-\frac{M}{r}\,N^2
   -2\,\frac{{\rm d}M}{{\rm d}\tau}\,\frac{{\rm d}r}{{\rm d}\tau}
   \pm M_{,u} \,.
\eeq

\subsection{Purely azimuthal motion ($\tilde\sigma\neq0$, $U^r=0$)}
\label{purely-azimuthal}

The second type of ``symmetric" motion is that along ``circular" orbits, namely with $U^r=0$. This only holds permanently (in order not to speak just of a turning point of the radial motion) if $U^r=0$ makes the right hand side of (\ref{Ur-contra-shorter}) zero,
\beq  \label{Ur-contra=0}
\fl
  \left( \frac{{\rm d}U^r}{{\rm d}\tau} =\right)
    -\frac{M}{r^2}+(r-3M)(U^\phi)^2
    \pm\frac{M_{,u}}{r}\,(U^u)^2
    \pm\tilde\sigma T_{uu}U^u
  =0 \,.
\eeq
The ``azimuthal" Eq.~(\ref{Uphi-contra}) yields, with $U^r=0$,
\beq
  \frac{{\rm d}U^\phi}{{\rm d}\tau}
  =-\tilde\sigma T_{uu}(U^u)^2 U^\phi
  =\pm\frac{\tilde\sigma M_{,u}}{4\pi}\,r^2(U^u)^2 U^\phi \,.
\eeq
Since the coefficient of $U^\phi $ is always negative if nonzero,
this means that if the particle interacts with the radiation, $U^\phi$ is slowed down to zero, which is the radiation drag effect.
Circular orbits with $U^\phi\neq 0$ thus do not seem to be possible, because the above condition $({{\rm d}U^r}/{{\rm d}\tau})|_{U^r=0}=0$ can hardly be constantly satisfied if $U^\phi$ is variable,
unless its variability were
exactly compensated by the effect of the change in $U^u$. This would require a very special adjustment of the $M$, $U^\phi$ and $U^u$ evolutions (for given values of $r={\rm const}$ and $\tilde\sigma$). Substituting for $T_{uu}$ and for $(U^\phi)^2$ from the normalization (\ref{norm-contra}), the condition (\ref{Ur-contra=0}) for $r={\rm const}$ becomes
\begin{eqnarray}
\fl
  0\left(=\frac{{\rm d}U^r}{{\rm d}\tau}\right)
   &=&-\frac{M}{r^2}+(r-3M)(U^\phi)^2
       \pm\frac{M_{,u}U^u}{r}\left(U^u\mp\frac{\tilde\sigma}{4\pi r}\right)
   \label{dUr/dtau=0} \\ {}
\fl
   &=&-\frac{M}{r^2}+\frac{r-3M}{r^2}\left[N^2(U^u)^2-1\right]
       \pm\frac{M_{,u}U^u}{r}\left(U^u\mp\frac{\tilde\sigma}{4\pi r}\right).
\end{eqnarray}
The positive solution of this last equation reads
\beq  \label{Uu-circular}
\fl
  U^u=\frac{r\tilde\sigma M_{,u}+
            \sqrt{(r\tilde\sigma M_{,u})^2+
                  64\pi^2 r(r-2M)\left[(r-2M)(r-3M)\pm r^2 M_{,u}\right]}}
           {8\pi\left[(r-2M)(r-3M)\pm r^2 M_{,u}\right]} \;,
\eeq
plus the normalization condition $r^2(U^\phi)^2=N^2(U^u)^2-1$ must  hold. (Remember that the upper signs correspond to $M_{,u}<0$ and the lower signs to $M_{,u}>0$.) These conditions ensure that at some instant of $u$, the test particle moves along $r={\rm const}$; the variation of $M$ with $u$ however makes them $u$-dependent, so the respective values of $U^u$, $U^\phi$ and $r$ change with $u$ as well.

It is also possible to use the normalization in the remaining time equation (\ref{Uu-contra}) to obtain
\beq
  \frac{{\rm d}U^u}{{\rm d}\tau}=
    \frac{\pm1}{N^2}\left[\frac{M}{r^2}-(r-3M)(U^\phi)^2\right]
    -\tilde\sigma T_{uu}(U^u)^3 \,,
\eeq
where one can in turn substitute from the above ${{\rm d}U^r}/{{\rm d}\tau}=0$ condition and rewrite the result in terms of the $u$-dependence (divide by $U^u$),
\begin{eqnarray}
  \frac{{\rm d}U^u}{{\rm d}u}
  &=&\frac{M_{,u}}{rN^2}
     \left[U^u\pm\frac{\tilde\sigma r}{4\pi}\,(U^\phi)^2\right] \nonumber \\
  &=&\frac{M_{,u}}{rN^2}
     \left\{U^u\pm\frac{\tilde\sigma}{4\pi r}\,[N^2(U^u)^2-1]\right\}\,.
  \label{dUu/du,circular}
\end{eqnarray}

In the geodesic limit ($\tilde\sigma=0$) the solution (\ref{Uu-circular}) and respective $U^\phi$ reduce to
\begin{eqnarray}
  (U^u)^2&=&\frac{r(r-2M)}{(r-2M)(r-3M)\pm r^2 M_{,u}} \,,
    \label{U^u-circular-geod} \\
  (U^\phi)^2&=&\frac{1}{r^2}\,
               \frac{M(r-2M)\mp r^2 M_{,u}}{(r-2M)(r-3M)\pm r^2 M_{,u}} \label{U^phi-circular-geod}\,.
\end{eqnarray}
The latter has to equal $\tilde{L}^2/r^4$, which yields
\begin{eqnarray}
\fl
  \lefteqn{
  \pm r^2(r^2+\tilde{L}^2)M_{,u}-(r-2M)\left[Mr^2-\tilde{L}^2(r-3M)\right]} \nonumber\\
\fl
  \lefteqn{
  =\pm M_{,u}r^4-Mr^3+(2M^2+\tilde{L}^2\pm\tilde{L}^2 M_{,u})r^2-5M\tilde{L}^2 r+6M^2\tilde{L}^2
  =0 \;.}  \label{circularity-timelike}
\end{eqnarray}
This can be understood either as a quartic equation for $r$, or as a ``compatibility condition" for $M_{,u}\,$, with solution
\beq  \label{M,u-circular-geo}
  \pm M_{,u}=\frac{(r-2M)\left[Mr^2-\tilde{L}^2(r-3M)\right]}{r^2(r^2+\tilde{L}^2)} \;.
\eeq
The left hand side $\pm M_{,u}$ is never positive, so the condition is only consistent if $\tilde{L}^2(r-3M)>Mr^2$. In such a case, its integration gives
\beq  \label{M(u)-compatibility}
\fl\quad
  M(u)=r\,\frac{(r-2M_0)\tilde{L}^2+\left[M_0 r^2-\tilde{L}^2(r-3M_0)\right]e^{\pm u/r}}
               {(r-2M_0)(r^2+3\tilde{L}^2)+2\left[M_0 r^2-\tilde{L}^2(r-3M_0)\right]e^{\pm u/r}}
  \,,
\eeq
where the integration constant has been chosen so that $M(0)=M_0$. Thus one must  have a very particular complicated mass function which depends on the special orbit parameters (radius and angular momentum) to allow circular geodesic orbits to exist.

For such a special orbit, knowing that $(U^\phi)^2$ must equal $\tilde{L}^2/r^4$, one can of course express $(U^u)^2$ in terms of $\tilde{L}^2$ from the normalization, as an alternative to (\ref{U^u-circular-geod}):
\beq
  (U^u)^2 =\frac{r^2+\tilde{L}^2}{r^2 N^2} \,.
\eeq
The square of the energy of the particle on this circular geodesic is then
\beq
\fl\quad
  (-U_u)^2=(-g_{uu}U^u)^2
          =\frac{(r-2M)^3}{r(r-2M)(r-3M)\pm r^3 M_{,u}}
          =N^2\left(1+\frac{\tilde{L}^2}{r^2}\right) \;.
\eeq

\subsection{Quasi-equilibrium locations ($U^i=0$)}
\label{equilibrium-locations}

Finally, it is natural to ask whether it is possible for the particle to remain with $U^i=0$ at some radii. Staying in equilibrium is a limiting case of purely radial motion (if in addition $U^r=0$) as well as of purely azimuthal motion (if in addition $U^\phi=0$). The normalization condition (\ref{norm-contra}) then implies
\beq
  U^u=N^{-1} \;,
\eeq
the only other constraint being the fulfillment of (\ref{dUr/dtau=0}), namely compensation of the radial gravitational pull, modified by the change of the central mass $M$, by the force exerted on the particle through the radiative flux,
\beq  \label{equilibrium-equation}
  -\frac{M}{r^2}
  \pm\frac{M_{,u}U^u}{r}\left(U^u\mp\frac{\tilde\sigma}{4\pi r}\right)=0 \,.
\eeq
Substituting for $U^u$ from above leads to
\beq  \label{equilibrium-equation,explicit}
  -M  \pm\frac{M_{,u}r}{N^2}
  -\frac{\tilde\sigma M_{,u}}{4\pi N}=0 \,.
\eeq
This is a quartic equation for $r$ which depends on the function $M$ of $u$, which means if $M$ is not constant, there is no true equilibrium in general, but in the quasi-stationary case in which $M$ changes sufficiently slowly, the roots of this equation will also change sufficiently slowly so as to be called quasi-equilibrium radii that will certainly influence the qualitative behavior of the orbits in a way similar to the actual equilibrium radii in the stationary case in which $M$ is constant.
However, one can also express the solution of this condition in terms of $M_{,u}$:
\beq  \label{equilibrium}
  \pm M_{,u}=\frac{MN^2}{r\mp\frac{\tilde\sigma}{4\pi}\,N} \;.
\eeq
This equation also gives the condition for a particular limiting case: the quasi-equilibrium can only be ``permanent," namely the corresponding radius remains at a given value $r$ for every $u$, if $M$ evolves with $u$ so that the condition is constantly satisfied for that fixed value of $r$. Clearly the condition can only hold with the upper sign, thus for $M_{,u}<0$ (outgoing flux) --- namely when the interaction term with $\tilde\sigma$ is larger than the first term $r$ in the denominator to make the right hand side negative. A necessary condition for equilibrium to be possible at some given fixed $r$ is therefore
\beq  \label{necessary-condition}
  \tilde\sigma>\frac{4\pi r}{N} \,.
\eeq
The solution of (\ref{equilibrium}) can be written implicitly as 
\beq
  u\pm r\ln\left(\frac{r-2M}{M}\right)
  -\frac{\tilde\sigma}{2\pi}\;{\rm arctanh}\,N
  ={\rm const} \;.
\eeq

``Switching off" the physical radiation-particle interaction ($\tilde\sigma=0$), one finds the condition for geodesic equilibrium,
\beq
  \pm M_{,u}=\frac{M}{r}\,N^2 \;\;(\,>0\,)\,.
\eeq
This condition can never be satisfied, because the left hand side is non-positive. This is due to the fact that the change of mass always makes the combined gravitational pull $-M/r^2 \pm{M_{,u}}/({rN^2})$ stronger (more negative) than the first term alone, irrespectively of the sign of $M_{,u}\,$ (see Eq.~(\ref{Ur-contra-shorter}) with $U^\phi=0$ and $\tilde\sigma=0$).

\section{Equilibrium solutions and the Schwarzschild limit with test radiation}
\label{Schw-comparison}

Let us discuss the special, lower-dimensional types of motion of the previous section in more detail and compare the Vaidya results with those obtained in a Schwarzschild background with a test radiation flux. 
The test-flux case has been treated both in the Schwarzschild \cite{BiniJS-09} and Kerr \cite{BiniGJSS-11} backgrounds
as a relativistic generalization of the classic Poynting-Robertson effect. The test particles moving in these spacetimes were subjected to a Thomson-type interaction with a superimposed test radiation field,
hence their motion was described by Eq.~(\ref{motion}) with $T^{\mu\nu}=\Phi^2 k^\mu k^\nu$ as in the present paper, but this null dust was taken there to be a test field. In contrast, in the present discussion the radiation is self-consistent,  being a source in the Vaidya exact solution of the Einstein equations. Since general test motion has to be solved numerically in both cases, we are mainly interested in the comparison of conditions for special types of motion which were obtained analytically. In particular, we will focus on the possibility of equilibrium locations, as noted in section \ref{equilibrium-locations}, or perhaps more appropriately in the present case, quasi-equilibrium locations.

In the Schwarzschild spacetime with purely radial test flux, the equilibrium condition takes the form (see Eq.~(3.22) of \cite{BiniJS-09})
\beq  \label{A/M}
  \frac{A}{M}=\pm N \;,
  \;\;\;\;\;\; {\rm where} \;\;\;\;
  A=\tilde\sigma\sqrt{g_{\theta\theta}g_{\phi\phi}}\;\Phi^2(-k_u)^2 \;.
\eeq
(We have added the $\pm$ sign here in order to include both the outgoing- and ingoing-radiation case, but it is immediately clear that it can only hold with outgoing radiation, i.e., with the plus sign.)
To verify whether this form of the equilibrium condition is consistent with the condition (\ref{equilibrium}) we derived above for the Vaidya field, we must translate the notation of that article. Substituting into (\ref{A/M})
\[ \sqrt{g_{\theta\theta}g_{\phi\phi}}|_{\theta =\pi/2}=r^2, \;\;\;\;
  \Phi^2=2T_{uu}=\mp\frac{M_{,u}}{2\pi r^2} \,, \;\;\;\;
  (-k_u)^2=\frac{1}{2} \,,\]
and noting that $A=\tilde\sigma T_{uu}r^2=\mp{\tilde\sigma M_{,u}}/({4\pi})$, the equilibrium condition becomes
\beq  \label{A/M-explicit}
  \mp\frac{\tilde\sigma M_{,u}}{4\pi M}=\pm N
  \;\;\;\; \Longleftrightarrow \;\;\;\;
  M_{,u}=-\frac{4\pi M}{\tilde\sigma}\,N \;.
\eeq
This coincides exactly with the condition following from Eq.~(\ref{equilibrium-equation}) if the term describing the gravitational effect of mass change (namely the one proportional to $M_{,u}$ but not containing $\tilde\sigma$) is omitted. Actually, this term corresponds exactly to the $r$ term in the denominator of (\ref{equilibrium}), so without it that condition reduces exactly to (\ref{A/M-explicit}). Expressed the other way round, switching from the Schwarzschild spacetime with a test radiation flux to the exact Vaidya spacetime with a self-consistent flux brings that $r$ term into the equilibrium condition (\ref{equilibrium}). This term makes the necessary equilibrium $M_{,u}$ more negative, namely bigger in absolute value. In order to understand this, it is convenient to look once more at the equilibrium condition (\ref{equilibrium-equation,explicit}),
\beq
  -\frac{M}{r^2}
  +\frac{M_{,u}}{r}\,\frac{1}{1-\frac{2M}{r}}
  -\frac{\tilde\sigma}{4\pi}\frac{M_{,u}}{r^2}\frac{1}{\sqrt{1-\frac{2M}{r}}}
  =0
\eeq
(limited to the upper sign since equilibrium is only possible in that case).
It is seen that the second term, describing the gravitational effect of mass change $M_{,u}\,$, falls off more slowly than the third term which describes physical effect of the flux corresponding to that $M_{,u}$ (and also more slowly than the first, Schwarzschild term). Therefore, it is difficult to reach equilibrium at very large radii, because the second term dominates there and the third term can only balance it with very large value of $\tilde\sigma$.

It is very useful to express the solution of the above equilibrium equation as
\beq  \label{equilibrium-in-r}
  r=N\left(\frac{\tilde\sigma}{4\pi}-\frac{MN}{|M_{,u}|}\right).
\eeq
It is seen from here that if the central mass were to turn completely into radiation ($M\rightarrow 0^+$, thus $N\rightarrow 1^-$), there would be two possibilities for the quasi-equilibrium radius: either it decreases to zero together with the mass, or it approaches the value
\beq
  r\rightarrow \frac{\tilde\sigma}{4\pi}-\frac{M}{|M_{,u}|}\equiv r_{\rm fin} \,.
\eeq
In particular, if $M/M_{,u}$ vanished in this limit, $r_{\rm fin}$ would be $\tilde\sigma/(4\pi)$ and would surely represent the maximum reached during the entire evolution of $r(u)$, because regardless of how $M$ and $r$ would behave precisely, one has (for $r>0$ of course) $0\leq N\leq 1$ and so
\[r=N\left(\frac{\tilde\sigma}{4\pi}-\frac{MN}{|M_{,u}|}\right)
   \leq N\,\frac{\tilde\sigma}{4\pi}
   \leq \frac{\tilde\sigma}{4\pi} \,.\]
It is also seen from (\ref{equilibrium-in-r}) that when $M_{,u}$ is so large that the second term is negligible, the quasi-equilibrium radius evolves according to $r\approx N\tilde\sigma/(4\pi)$.

\subsection*{Specific choices for $M(u)$}

To illustrate properties of the equilibrium condition, we choose a few particular mass functions $M(u)$. We consider the case where $A/M={\rm const}$ (which implies exponential decay of $M$), the case where the mass $M$ decreases linearly, and the case of a hyperbolic tangent mass profile often used in the literature (see, e.g., Ref.~\cite{singh}).

\subsubsection*{$\bullet$
$A/M={\rm const}$: exponential decrease of $M$.}

%%%%%%%%%%%%%%%%%%%%%%%%%%%%%%%%%%%%%%%%%%%%%%%%%%%%%%%%%%%%%%%%%
\begin{figure}
\begin{center}
\includegraphics[scale=0.55]{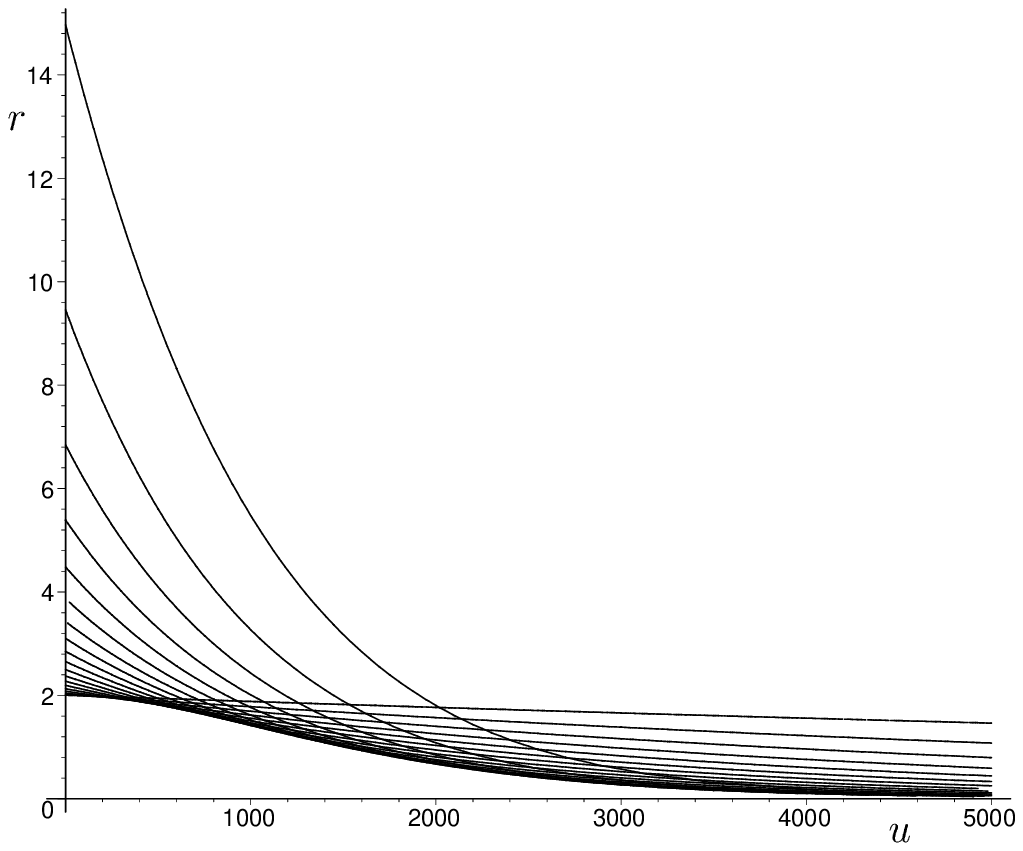}
\end{center}
\caption
{Evolution of the quasi-equilibrium radius $r$ with time $u$ for exponential mass decrease $M(u)=M_0\exp(-4\pi u\A/\tilde\sigma)$ (which yields $\A={\rm const}$) with $\tilde\sigma=10^4$, as given by (\ref{equilibrium-in-r,expo}). The curves (going from top to bottom along the vertical axis)  correspond to $\A=0.95$, 0.9, 0.85, 0.8, \dots, 0.1, 0.05.
The axes are in the units of $M_0$. (The decrease of $r$ below $2M_0$ does not mean that it is below the apparent horizon radius, because the actual $M$ also decreases.)
}
\label{ru-expo}
\end{figure}
%%%%%%%%%%%%%%%%%%%%%%%%%%%%%%%%%%%%%%%%%%%%%%%%%%%%%%%%%%%%%%%%%

\null

Let us choose $A/M$ to be some positive constant (call it $\A $), which corresponds to the case when the luminosity at infinity equals the constant fraction $\A $ of the Eddington value  (see section 3.2 of \cite{BiniJS-09} and Eq.~(2.33) in \cite{BiniGJSS-11}). The particular behavior of mass which ensures this, $M_{,u}=-4\pi M\A /\tilde\sigma$, namely exponential decay $M(u)=M_0\exp(-4\pi u\A /\tilde\sigma)$, makes the equilibrium condition yield
\beq  \label{equilibrium-in-r,expo}
  r=\frac{\tilde\sigma}{4\pi}\,\frac{N\,(\A -N)}{\A } \,.
\eeq
Note that in this case
\[\frac{M}{|M_{,u}|}=\frac{\tilde\sigma}{4\pi\A }\neq 0 \,,\]
hence the ultimate value of the quasi-equilibrium radius (reached at $u\rightarrow\infty$ here) is certainly not $\tilde\sigma/(4\pi)$. Its behavior at final stages of the center's ``evaporation" can be inferred by linearizing the equilibrium condition (and then again its solution $r$) in $M$ which yields
\[r(u\rightarrow\infty)=\frac{2-\A }{1-\A }\,M \,.\]
Hence the quasi-equilibrium radius tends to zero exponentially together with $M$.
Fig.~\ref{ru-expo} shows an example of the dependence of the equilibrium curves $r(u;\A )$ on the parameter $\A $ for a given $\tilde\sigma$.

\subsubsection*{$\bullet$
Linear decrease of $M$.}

%%%%%%%%%%%%%%%%%%%%%%%%%%%%%%%%%%%%%%%%%%%%%%%%%%%%%%%%%%%%%%%%%
\begin{figure}
\begin{center}
$\begin{array}{cc}
\includegraphics[scale=0.55]{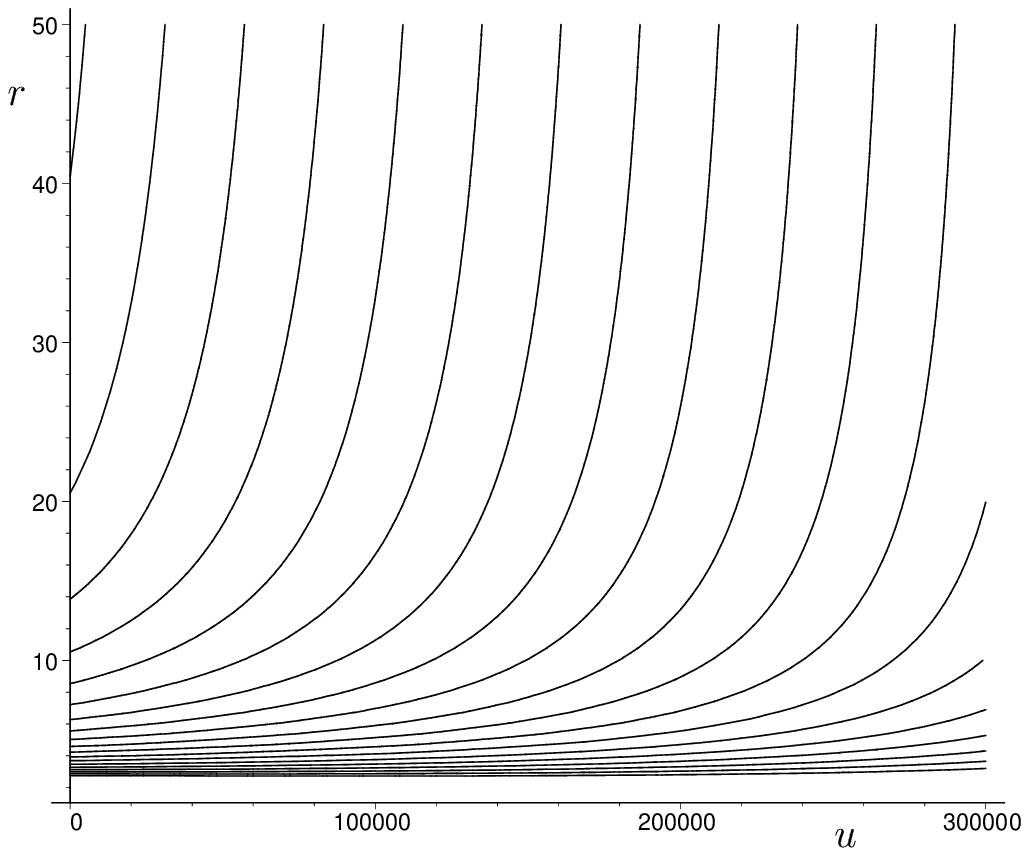}&\quad
\includegraphics[scale=0.55]{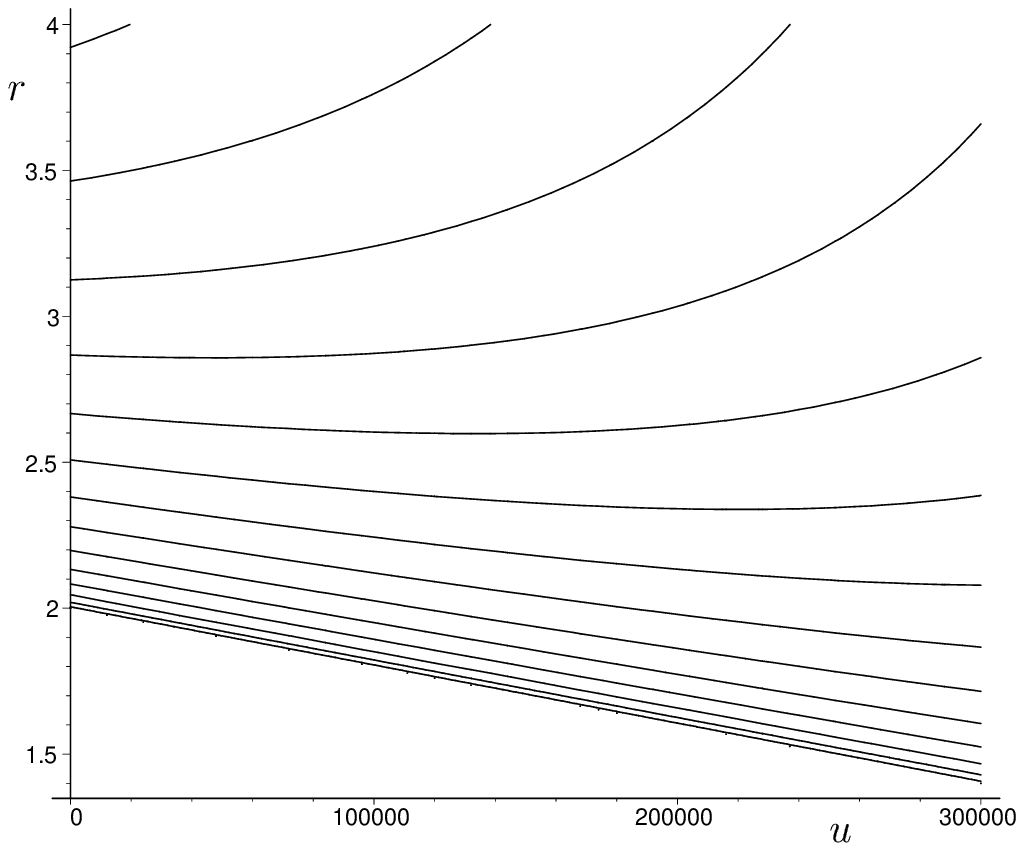}\\[.4cm]
\quad\mbox{(a)}\quad &\quad \mbox{(b)}
\end{array}$\\
\end{center}
\caption
{Evolution of the quasi-equilibrium radius $r$ with time $u$ for linear mass decrease $M(u)=M_0-\beta u$ with $\beta=10^{-6}$, as given by (\ref{equilibrium-in-r,linear}); 3/10 of the center's lifetime $M_0/\beta$ are covered.
On the plot (a), the curves correspond (going from top to bottom) to $\A _0=0.975$, 0.95, 0.925, 0.9, \dots, 0.55, 0.525.
On the plot (b), the curves obtained for smaller values of $\A _0$, namely (from top to bottom) 0.7, 0.65, 0.6, 0.55, \dots, 0.1, 0.05, are shown in more detail.
The axes are in the units of $M_0$. (The decrease of $r$ below $2M_0$ does not mean that it is below the apparent horizon, because the actual $M$ also decreases.)}
\label{ru-linear}
\end{figure}
%%%%%%%%%%%%%%%%%%%%%%%%%%%%%%%%%%%%%%%%%%%%%%%%%%%%%%%%%%%%%%%%%

Another natural possibility is a linear decrease of mass, $M(u)=M_0-\beta u$. In such a case the equilibrium condition (\ref{equilibrium-in-r}) reads
\beq  \label{equilibrium-in-r,linear}
  r=N\left(\frac{\tilde\sigma}{4\pi}-\frac{MN}{\beta}\right)
\eeq
and the ultimate value of the quasi-equilibrium radius is $r_{\rm fin}=\tilde\sigma/(4\pi)$. This radius is a global maximum of $r(u)$; actually $r(u)$ has an overall tendency to grow. This is expected since the linear decrease of mass corresponds to the increase of the effective interaction parameter $\frac{A}{M}=\frac{\tilde\sigma\beta}{4\pi M}\,$. The quasi-equilibrium radius in fact need not increase all the time (for small $\tilde\sigma$, it rather decreases initially and later remains more or less constant), but finally (at times $u\sim M_0/\beta$) it always grows with $u$. The rate of this growth is again obtained by restricting just to linear terms in $M$ in the above equation; this yields
\[r(u\rightarrow M_0/\beta)=r_{\rm fin}-\frac{1+\beta}{\beta}\,M \,.\]

The quasi-equilibrium $r(u)$ dependence is parameterized by $\beta$ and $\tilde\sigma$, $M_0$ playing the role of a scale factor. For illustration, one can either choose a certain fixed $\tilde\sigma$ and plot $r(u)$ for different possible $\beta$, but it it more suitable to do it the other way round, because the ``lifetime of the star" is given by $M_0/\beta$ and it is better to have the latter the same for all the curves. See Fig.~\ref{ru-linear} for an example of the dependence of such quasi-equilibrium curves $r(u;\A _0)$ on $\A _0\equiv\frac{A}{M_0}=\frac{\tilde \sigma \beta}{4\pi M_0}\,$, for a given $\beta$. In accord with formula (\ref{r-equilibrium-linear}) derived below, the quasi-equilibrium radius first drifts up/down if $\A _0 $ is bigger/smaller than $1/\sqrt{3}\,$.

Figs.~\ref{ru-expo} and \ref{ru-linear} show that the $r(u)$ behavior is rather different for the exponential and linear decrease of $M$ with $u$: in absolute measures (in units of the initial mass $M_0$), the exponential decrease of $M$ leads to decrease of $r$, whereas the linear decrease of $M$ leads to increase of $r$ with $u$. If expressed in units of the actual mass $M$, the quasi-equilibrium radii of course have stronger tendency to grow in time, and actually they (slightly) do so even in the exponential case. 
On a more general level, this reminds us that it is delicate (if at all possible) to compare locations in a non-stationary spacetime at different moments (and even harder to compare them in different spacetimes, i.e., with different $M(u)$).

\subsubsection*{$\bullet$
Hyperbolic tangent mass profile.}

%%%%%%%%%%%%%%%%%%%%%%%%%%%%%%%%%%%%%%%%%%%%%%%%%%%%%%%%%%%%%%%%%
\begin{figure}
\begin{center}
\includegraphics[scale=0.3]{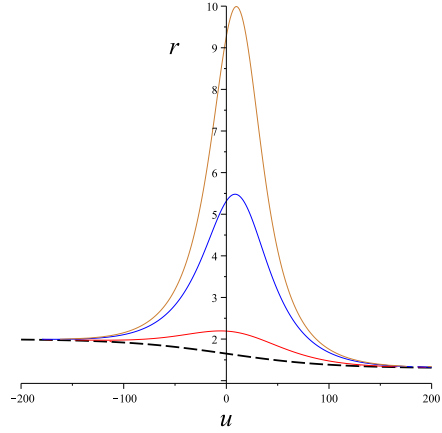}
\end{center}
\caption
{Evolution of the quasi-equilibrium radius $r$ with time $u$ for the mass profile $M(u)=M_1+(M_2-M_1)(1+\tanh\beta u)/2$ with the following parameter choice: $M_1=1$, $M_2=0.65$, $\beta=10^{-2}$ and different values of $\tilde\sigma=[3,5,5.5]\cdot 10^3$, with axes in units of $M_1$.
The black dashed curve corresponds to the apparent horizon.
During the transition phase the quasi-equilibrium radius is even more enhanced for increasing values of $\tilde\sigma$, i.e., when the interaction with the radiation field becomes stronger.
}
\label{ru-expo2}
\end{figure}
%%%%%%%%%%%%%%%%%%%%%%%%%%%%%%%%%%%%%%%%%%%%%%%%%%%%%%%%%%%%%%%%%

Finally, one can assume that the mass smoothly decreases/increases between some two fixed values $M_1$, $M_2$. 
This situation can be modeled by the mass profile 
\begin{eqnarray}\fl\quad
\label{massprofile}
  M(u)=M_1+\frac{(M_2-M_1)}{2}(1+\tanh\beta u)\,, 
\quad
  M_{,u}=\frac{\beta}{2} (M_2-M_1) \,{\rm sech}^2 \beta u
\end{eqnarray}
for the outgoing ($M_1>M_2$) and ingoing radiation ($M_1<M_2$) cases, where the constant rate parameter $\beta$ governs the time scale of the transition between the two asymptotic Schwarzschild spacetimes (the smaller the value of $\beta$ the longer the transition). 
The behavior of the quasi-equilibrium radius as a function of $u$ is shown in Fig.~\ref{ru-expo2} in the outgoing case for selected values of $\tilde\sigma$.
Interestingly, during the transition phase the quasi-equilibrium radius can reach large values when the interaction with the radiation field is very strong.

\subsection*{Approximate solution for very slow steady mass decay}
\label{limit}

To compare with the corresponding Schwarzschild test radiation case,
consider a linear mass decrease $M(u)=M_0-\beta u$ with constant and very small $\beta$ ($0\le \beta \ll 1$). 
The interaction parameter and its dimensionless counterpart are then
\beq
  A=\frac{\tilde \sigma \beta}{4\pi}
  \,,\qquad 
  \A_0\equiv\frac{A}{M_0} = \frac{\tilde \sigma \beta}{4\pi M_0} \;.
\eeq
Let us stress that $\A_0$ is not necessarily small (in contrast to $\beta$) since $\tilde\sigma$ can be large. Actually the necessary condition for the equilibrium to be at all possible, (\ref{necessary-condition}), requires
\beq
  \A_0>\frac{\beta r}{M_0 N} \,.
\eeq
By rewriting (\ref{equilibrium-in-r,linear}) as
\[\A_0=\frac{\beta r}{M_0 N}+N\frac{M}{M_0}\,,\]
one sees that the above condition can also be understood as the requirement of positivity of the mass $M$ (which is here ensured at times $u<M_0/\beta$).

Substituting for $M(u)$ in the lapse function yields
\[N^2=1-\frac{2M_0}{r}+\frac{2\beta u}{r}\equiv N_0^2+\frac{2\beta u}{r}\,,\]
so the equilibrium condition $MN^2+\beta r-M_0\A_0 N=0$ becomes
\beq
  \left(1-\beta\frac{u}{M_0}\right)\left(N_0^2+2\beta\,\frac{u}{r}\right)
  +\beta\frac{r}{M_0}-\A_0\,\sqrt{N_0^2+2\beta\,\frac{u}{r}}=0 \,.
\eeq
Expanding this now only up to terms linear in $\beta$, one obtains the dimensionless equation
\beq
  N_0(N_0-\A_0)+\beta
  \left[\frac{r}{M_0}-\frac{u}{M_0}
        \left(1-\frac{4M_0}{r}+\frac{\A_0}{N_0}\frac{M_0}{r}\right)\right]
  =0
\eeq
whose relevant solution (linearized in $\beta$) reads
\beq  \label{r-equilibrium-linear}
  r=\frac{2M_0}{1-\A^2}
    -\frac{2\beta}{(1-\A_0^2)^3}
     \left[4M_0+u\,(1-\A_0^2)(1-3\A_0^2)\right].
\eeq
In the limit when even the linear term can be neglected ($\beta\rightarrow 0$), we get the familiar Schwarzschild equilibrium solution for a test flux, $r=2M_0/(1-\A_0^2)\equiv r_0$, which only exists when $0\le \A_0<1$. Writing the above solution in dimensionless form ($\rho\equiv r/M$),
\beq
  \rho=\rho_0(1-\beta\rho_0^2)-\frac{\beta}{2}\,\rho_0^2\,(1-3\A _0^2)\,\frac{u}{M_0} \,,
\eeq
one sees that if $\beta$ cannot be neglected completely but the linear approximation is sufficient, the quasi-equilibrium radius starts from a slightly lower value $\rho=\rho_0(1-\beta\rho_0^2)$ than in the Schwarzschild case and drifts with time $u$ towards even smaller values when $\A_0<1/\sqrt{3}\approx 0.577$ whereas towards larger values when $\A_0>1/\sqrt{3}$; this initial behavior is clearly shown in Fig.~\ref{ru-linear}(b) in the zone before the total mass loss becomes comparable to the initial value and the curves quickly rise. During this initial phase, the quasi-equilibrium radius provides a slowly moving target near the corresponding Schwarzschild value.

Note that assuming the mass profile as in Eq.~(\ref{massprofile}) also leads to a constant rate of loss of energy that occurs in the test radiation field. In fact, for small values of $\beta$, i.e. for a very slow transition between a past asymptotic Schwarzschild spacetime with mass $M_1$ and a future asymptotic Schwarzschild spacetime with mass $M_2$, we have in the outgoing case $M_{,u}\sim-\beta(M_1-M_2)/2<0$, so that the same considerations as above apply too.

\subsection*{Condition for purely azimuthal motion}

One can also be interested in circular equilibrium orbits, i.e., with only $\nu^{\hat r}=0$.
In this case Eqs.~(\ref{motion2}) reduce to 
\begin{eqnarray}
\label{motion4uno}
0&=&
  -\frac{\gamma}{ rN^3}\left[\frac{M}{r}\,N^2\mp M_{,u}\right] +\frac{\gamma N}{r}(\nu^{\hat \phi}){}^2
  \pm\frac{\tilde \sigma T_{uu}}{N^2}\,, \\
  \label{motion4due}
\frac{\rmd \nu^{\hat \phi}}{\rmd \tau}&=&
 -\frac{\tilde \sigma T_{uu}}{N^2}\nu^{\hat \phi}\,.
\end{eqnarray}
Re-expressing $T_{uu}$ in terms of $M_{,\tau}$, the second equation can be solved for $\gamma=1/\sqrt{1-(\nu^{\hat \phi}){}^2}$ leading to
\beq
\gamma(\tau)=\frac{\gamma_0+\tanh\left[\frac{\pm\tilde\sigma (N(\tau)-N_0)}{4\pi r}\right]}{1+\gamma_0\tanh\left[\frac{\pm\tilde\sigma (N(\tau)-N_0)}{4\pi r}\right]}\,,
\eeq
where $\gamma_0=\gamma(0)$ and $N_0=N(0)$.
Solving then the latter equation for $\nu^{\hat \phi}$ and substituting into Eq. (\ref{motion4due}) gives a first order equation for $M(\tau)$.
Therefore, circular equilibrium exists only if $M$ evolves with $u$ according to this equation for $r=\,$ constant.
This condition selects a particular mass profile. 
Assuming instead a given mass profile, one obtains the evolution of the radius such that the orbit is momentarily circular, i.e., a quasi-equilibrium state.

\begin{figure} 
\begin{center}
$\begin{array}{cc}
\includegraphics[scale=0.3]{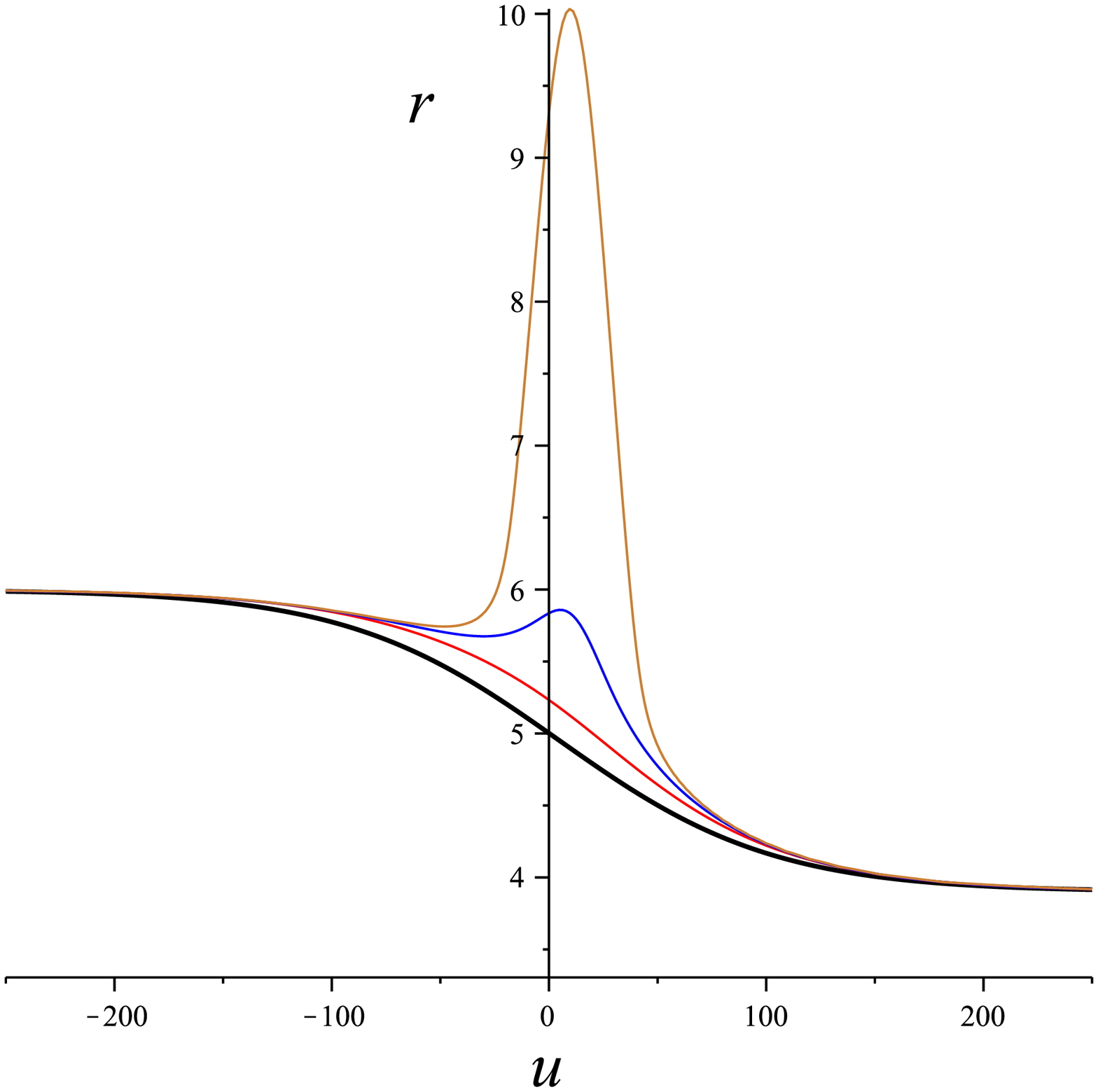}&\quad
\includegraphics[scale=0.3]{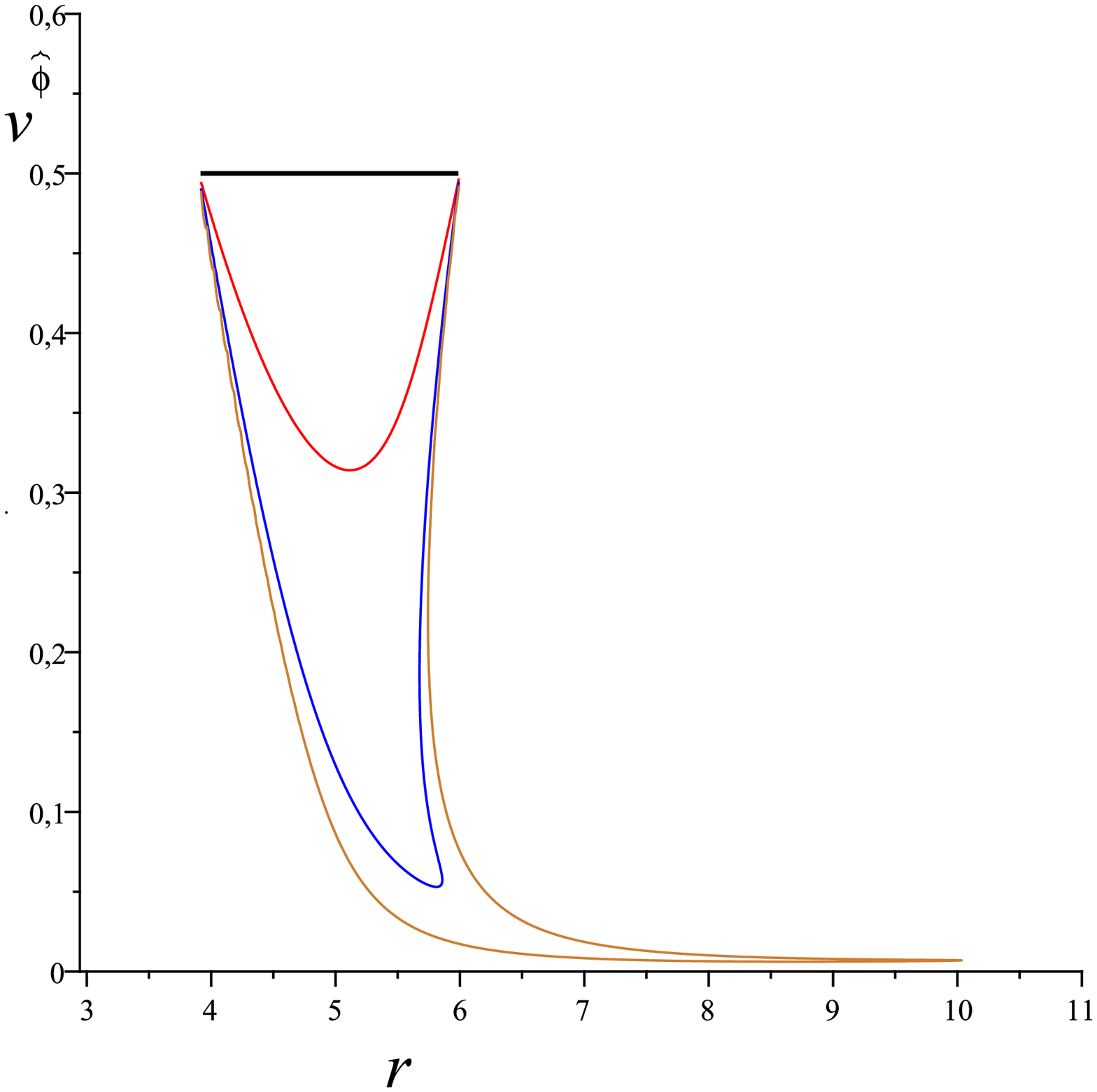}\\[.4cm]
\quad\mbox{(a)}\quad &\quad \mbox{(b)}\\
\end{array}$\\[.6cm]
\end{center}
\caption
{The evolution with time $u$ of the quasi-equilibrium radius $r$ corresponding to circular quasi-equilibrium orbits is shown in plot (a) for the mass profile $M(u)=M_1+(M_2-M_1)(1+\tanh\beta u)/2$ with the following parameter choice: $M_1=1$, $M_2=0.65$, $\beta=10^{-2}$ and selected values of $\tilde\sigma=[0,3,5,5.5]\cdot10^3$, with axes in units of $M_1$.
The initial value of the radius is taken to be $6M_1$.
The corresponding behavior of the azimuthal velocity $\nu^{\hat \phi}$ is shown in plot (b).
In the geodesic case (thick black curves) the constant value of the azimuthal velocity is set to $1/2$, which corresponds to the Keplerian value $\nu_K=\sqrt{M_1/(r_0-2M_1)}$ for $r_0=6M_1$ in the past asymptotic Schwarzschild spacetime.
The asymptotic value of the quasi-equilibrium radius after the transition is $6M_2=3.9M_1$.
}
\label{fig:rdiu_circ}
\end{figure}

Fig. \ref{fig:rdiu_circ} shows the evolution of such a quasi-equilibrium radius $r$ with time $u$ for the mass profile (\ref{massprofile}) in the outgoing case and the corresponding behavior of the azimuthal velocity $\nu^{\hat \phi}$.
In the case of geodesic motion $\nu^{\hat \phi}=\,$ constant and 
\beq
\label{nuphigeo}
(\nu^{\hat \phi}){}^2=\frac{M}{rN^2}-\frac{M_{,u}}{N^4}\,,
\eeq
according to Eqs. (\ref{motion4uno})--(\ref{motion4due}) (see also Eqs. (\ref{U^u-circular-geod})--(\ref{U^phi-circular-geod})), which is an equation for $r$.
The first term represents the Keplerian one.
Since in the outgoing case $M_{,u}$ is always negative, the second term in Eq.~(\ref{nuphigeo}) is always positive, so that the radius has to be greater than the Keplerian one.  
If the transition between the two asymptotic Schwarzschild spacetimes occurs very slowly, the second term is negligible so that the actual behavior of $r(u)$ is practically indistinguishable from the Keplerian one. 
The effect of the interaction with the radiation field is a growth of the quasi-equilibrium radius for purely azimuthal motion during the transition phase just as in the case of quasi-equilibrium at rest.

%%%%%%%%%%%%%%%%%%%%%%%%%%%%%%%%%%%%%%%%%%%%%%%%%%%%%%%%%%%%%%%%%
%%%%%%%%%%%%%%%%%%%%%%%%%%%%%%%%%%%%%%%%%%%%%%%%%%%%%%%%%%%%%%%%%

\section{Examples of numerical orbits}
\label{numerics}

\begin{figure} 
\begin{center}
$\begin{array}{cc}
\includegraphics[scale=0.3]{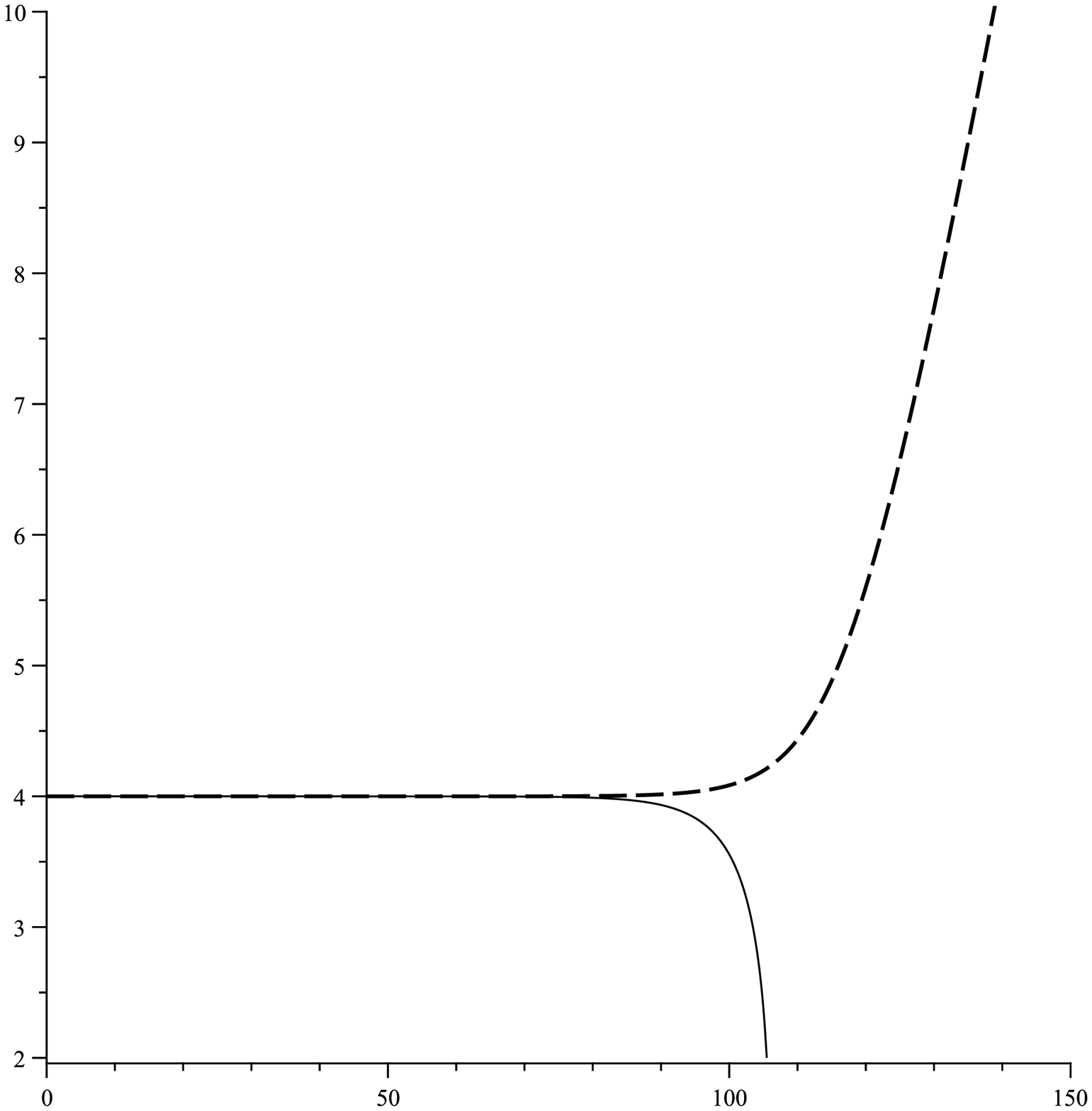}&\quad
\includegraphics[scale=0.3]{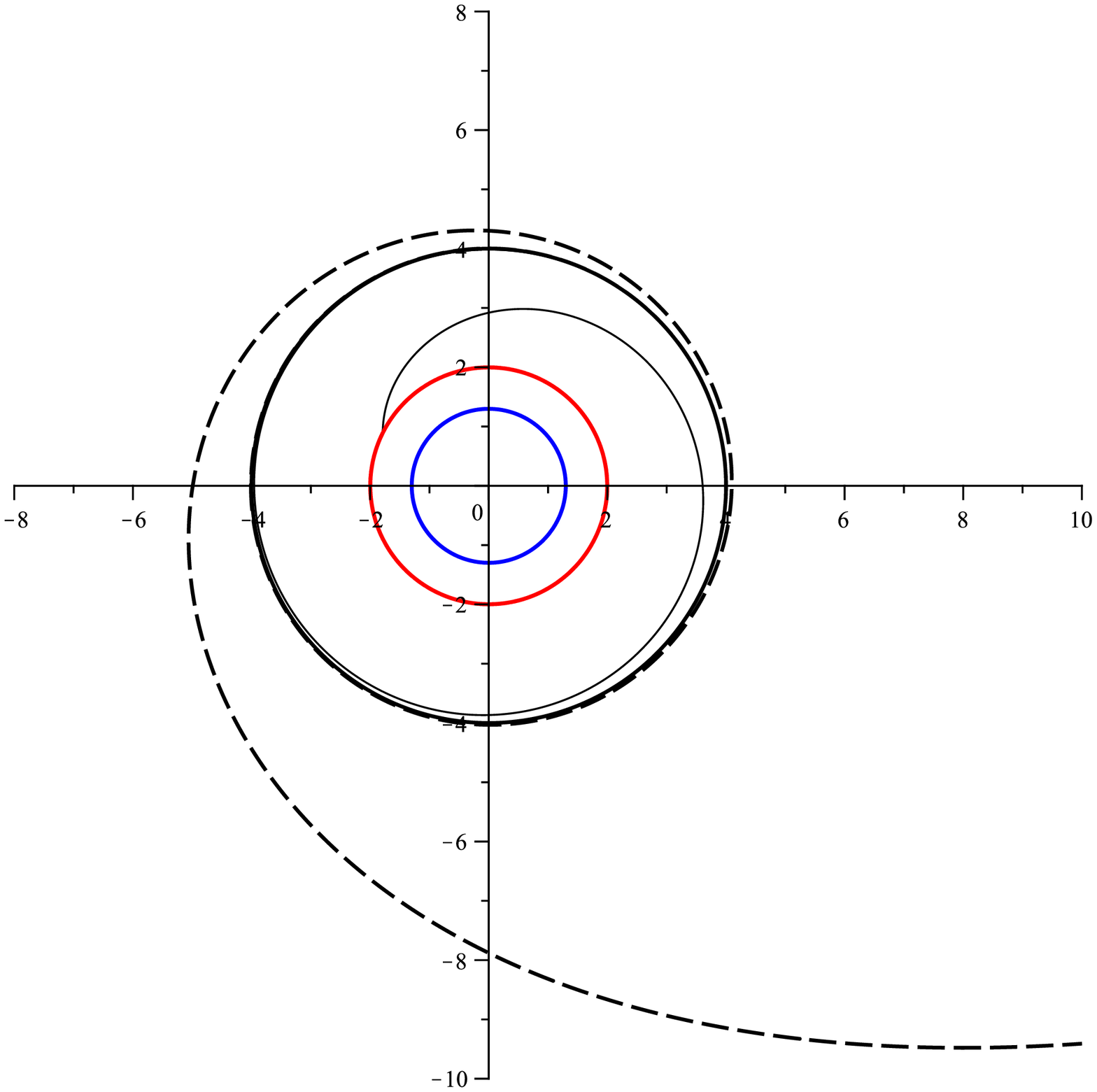}\\[.4cm]
\quad\mbox{(a)}\quad &\quad \mbox{(b)}
\end{array}$\\
\end{center}
\caption{The behavior of $r(\tau)$ is shown in (a) in the case of outgoing radiation with the following parameter choice: $M_1=1$, $M_2=0.65$, $\beta=10^{-2}$ and $\tilde\sigma=0$ (geodesic, thick dashed line) and $\tilde\sigma=10^4$ (solid line), with axes given in units of $M_1$.
The initial conditions are  $u(0)=-1000$, $r(0)=4$, $\phi(0)=0$, $\nu^{\hat r}(0)=0$, $\nu^{\hat \phi}(0)\approx0.707$, which correspond to a circular geodesic in the past asymptotic Schwarzschild spacetime with mass $M_1$. 
The corresponding orbits are shown in (b).
In the geodesic case, the orbit escapes outwards after a few loops.
In constrast, the accelerated particle spirals towards the apparent horizon, which is reached in a finite proper time interval at $r\approx2$.
The asymptotic inner apparent horizon at $r\approx1.3$ is also shown.
}
\label{fig:4}
\end{figure}

\begin{figure} 
\begin{center}
$\begin{array}{cc}
\includegraphics[scale=0.3]{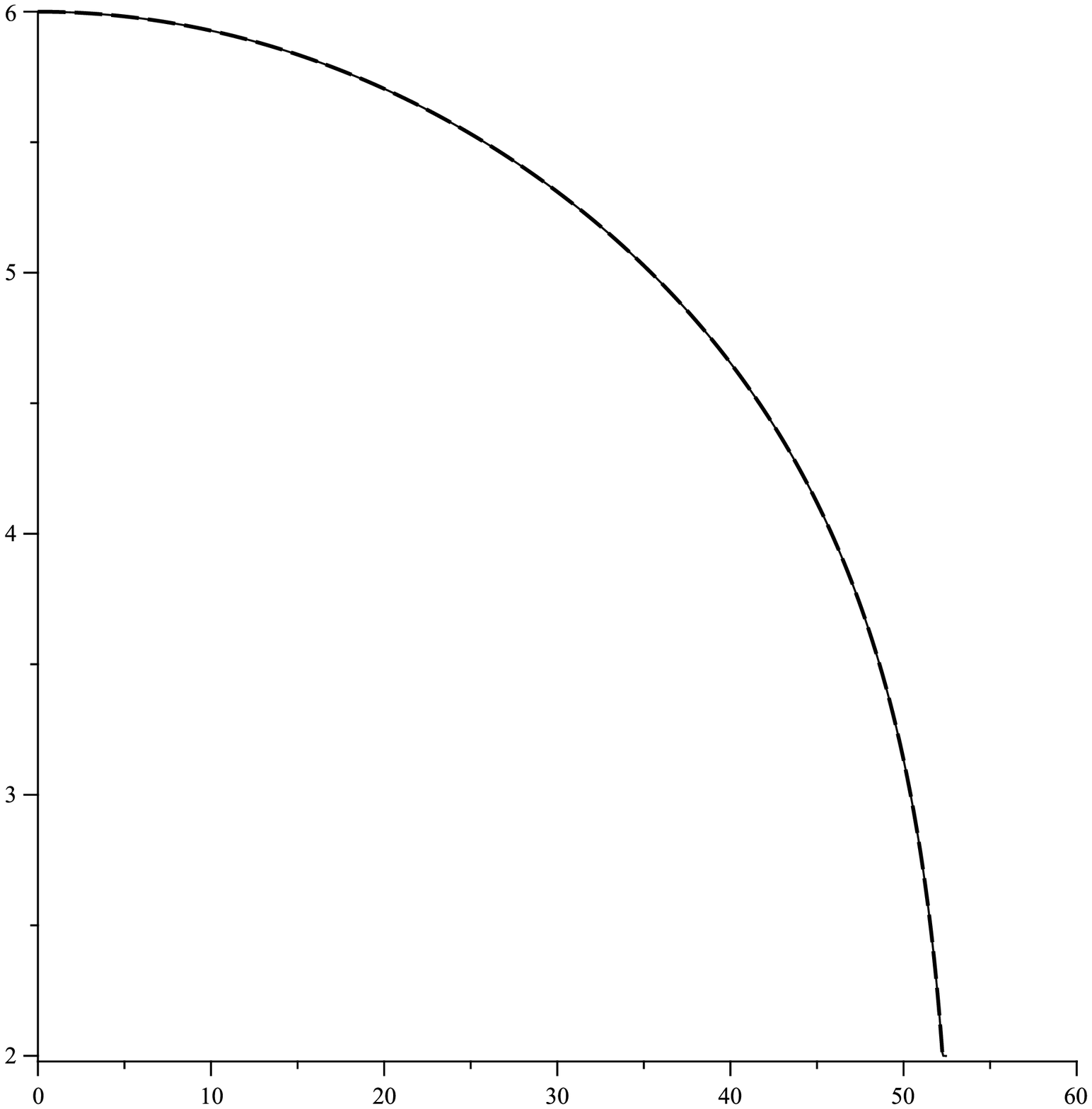}&\quad
\includegraphics[scale=0.3]{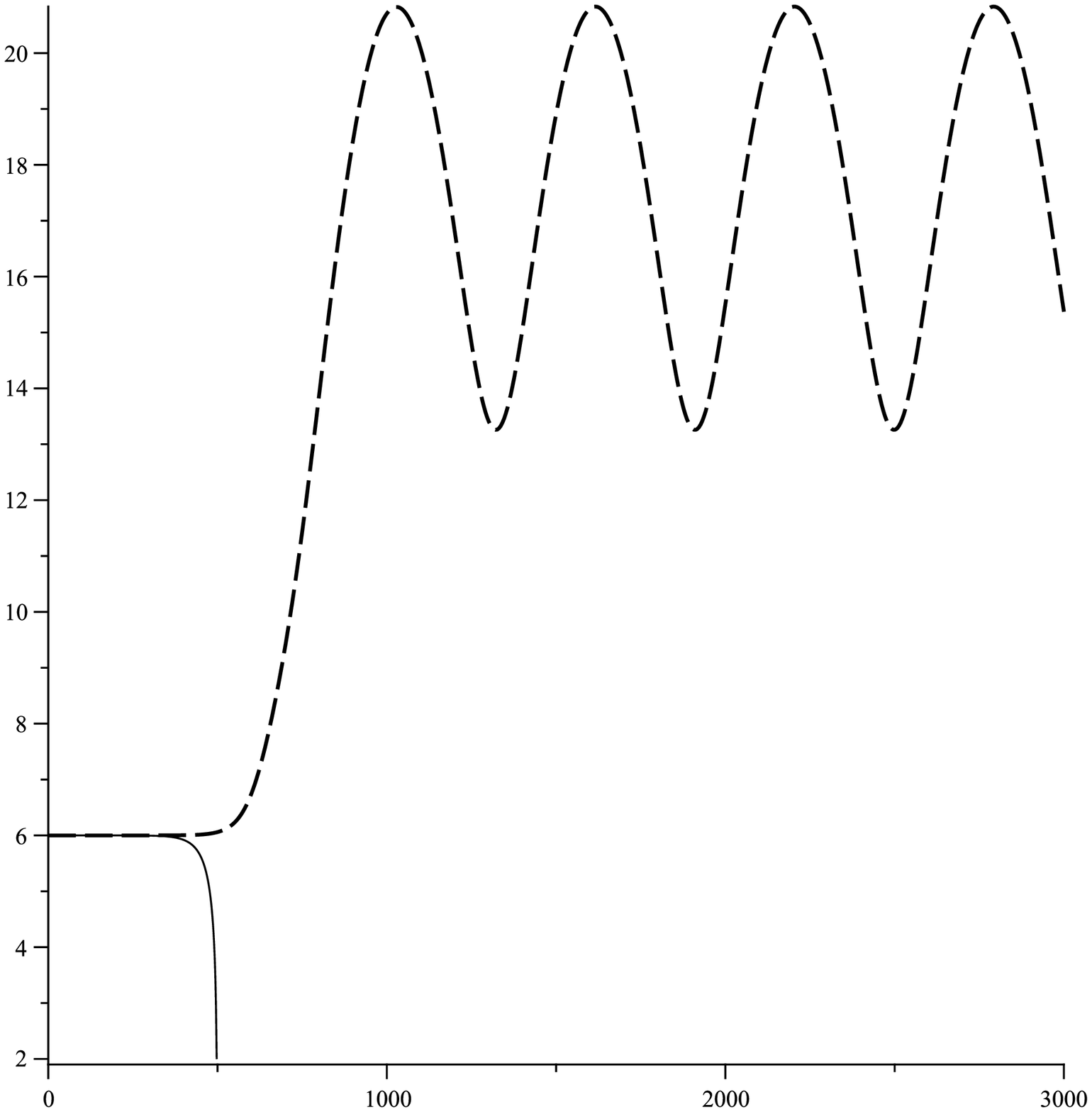}\\[.4cm]
\quad\mbox{(a)}\quad &\quad \mbox{(b)}\\
\end{array}$\\[.6cm]
\includegraphics[scale=0.3]{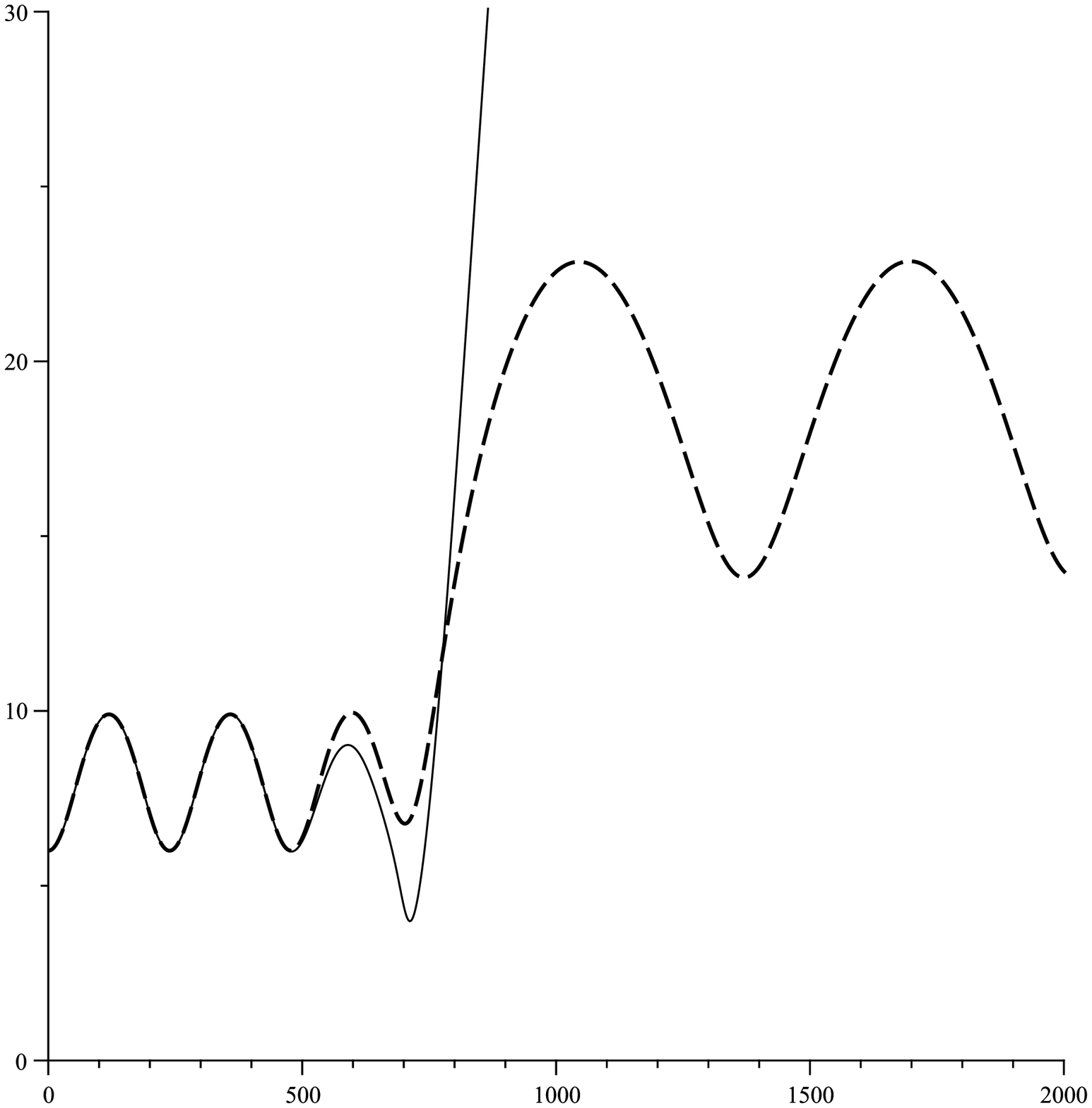}\\[.4cm]
\mbox{(c)}
\end{center}
\caption{The behavior of $r(\tau)$ is shown in the case of outgoing radiation with the same parameter choice as in Fig.~\ref{fig:4}, with axes given in units of $M_1$.
The initial conditions are $u(0)=-1000$, $r(0)=6$, $\phi(0)=0$, $\nu^{\hat r}(0)=0$ and $\nu^{\hat \phi}(0)=[0.49,0.5,0.51]$, in (a) to (c) respectively. 
The value $\nu^{\hat \phi}(0)=0.5$ corresponds to a circular geodesic in the past asymptotic Schwarzschild spacetime with mass $M_1$.
In this case shown in (b), the geodesic orbit exhibits an oscillating behavior while approaching the circular geodesic of the future asymptotic Schwarzschild spacetime with mass $M_2$ at Keplerian radius $r\approx17$ and Keplerian speed $\nu_K=\sqrt{M_2/(r-2M_2)}\approx0.2$.
In contrast, the accelerated particle spirals towards the apparent horizon  at $r\approx2$ and reaches the latter in a finite proper time interval.
Before falling into the apparent horizon, the value $\nu^{\hat \phi}(0)=0.49$ (and smaller values) corresponds to a spiraling behavior towards the apparent horizon in both the geodesic and accelerated cases.
The value $\nu^{\hat \phi}(0)=0.51$ shows oscillations outside the initial radius, from which the accelerated particles then escape.
Further increase of $\nu^{\hat \phi}(0)$ would lead to particle escape in both cases.
}
\label{fig:6}
\end{figure}

\begin{figure} 
\begin{center}
\includegraphics[scale=0.3]{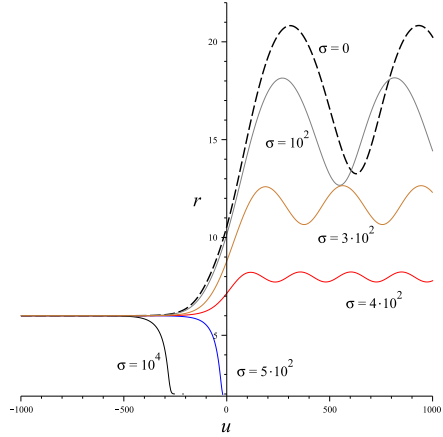}
\end{center}
\caption{The behavior of $r(u)$ is shown in the case of outgoing radiation with the same parameter choice and initial conditions as in Fig.~\ref{fig:6}(b), but for different values of the friction parameter $\tilde\sigma$.
For very small values of $\tilde\sigma$ the accelerated orbit is close to the geodesic one.
As the interaction with the background radiation field becomes stronger, i.e., for increasing values of $\tilde\sigma$, the asymptotic radius decreases to even smaller values.
Further increasing of $\tilde\sigma$ causes the particle to cross the apparent horizon.
}
\label{fig:7}
\end{figure}

\begin{figure} 
\begin{center}
$\begin{array}{cc}
\includegraphics[scale=0.3]{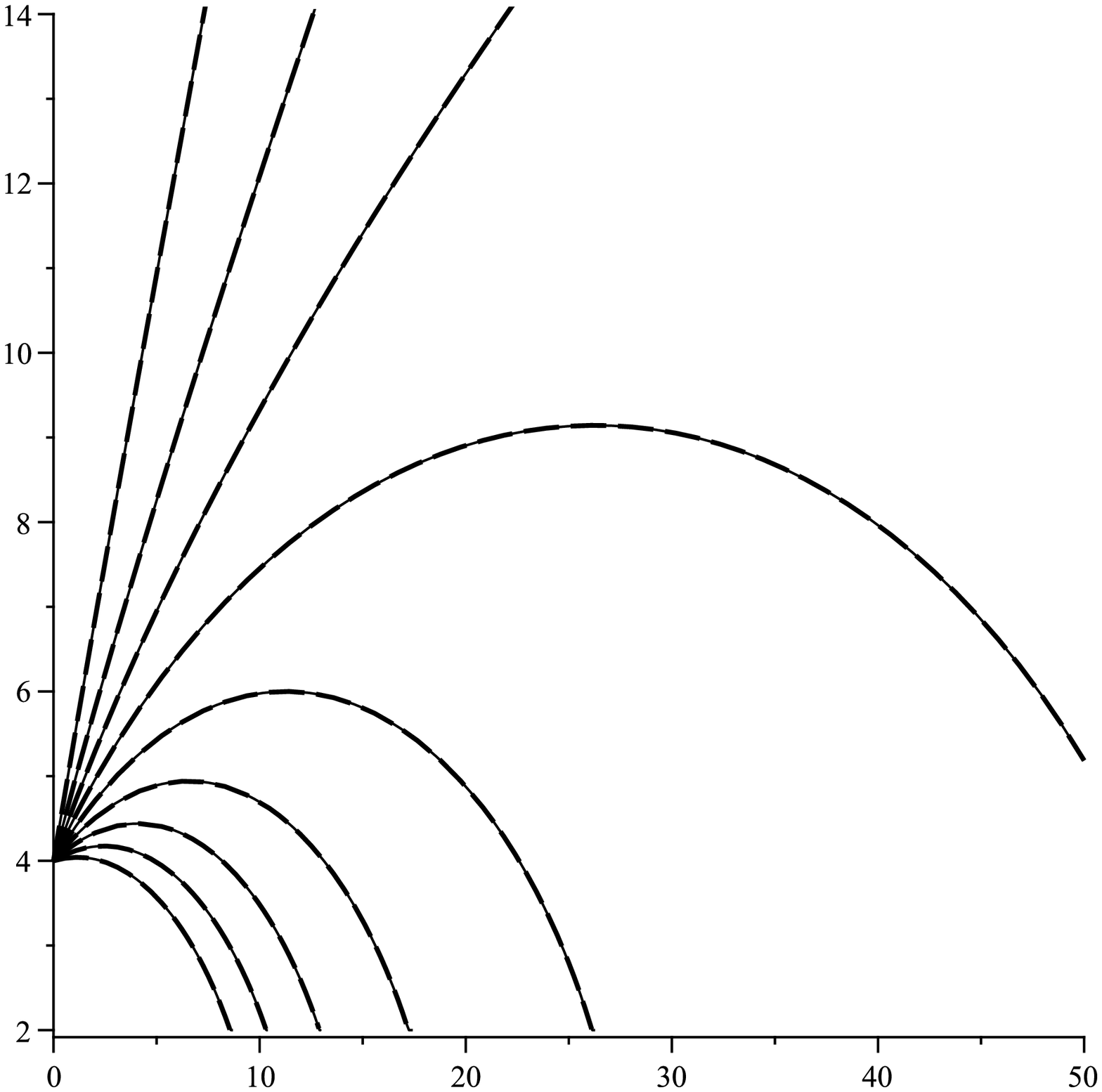}&\quad
\includegraphics[scale=0.3]{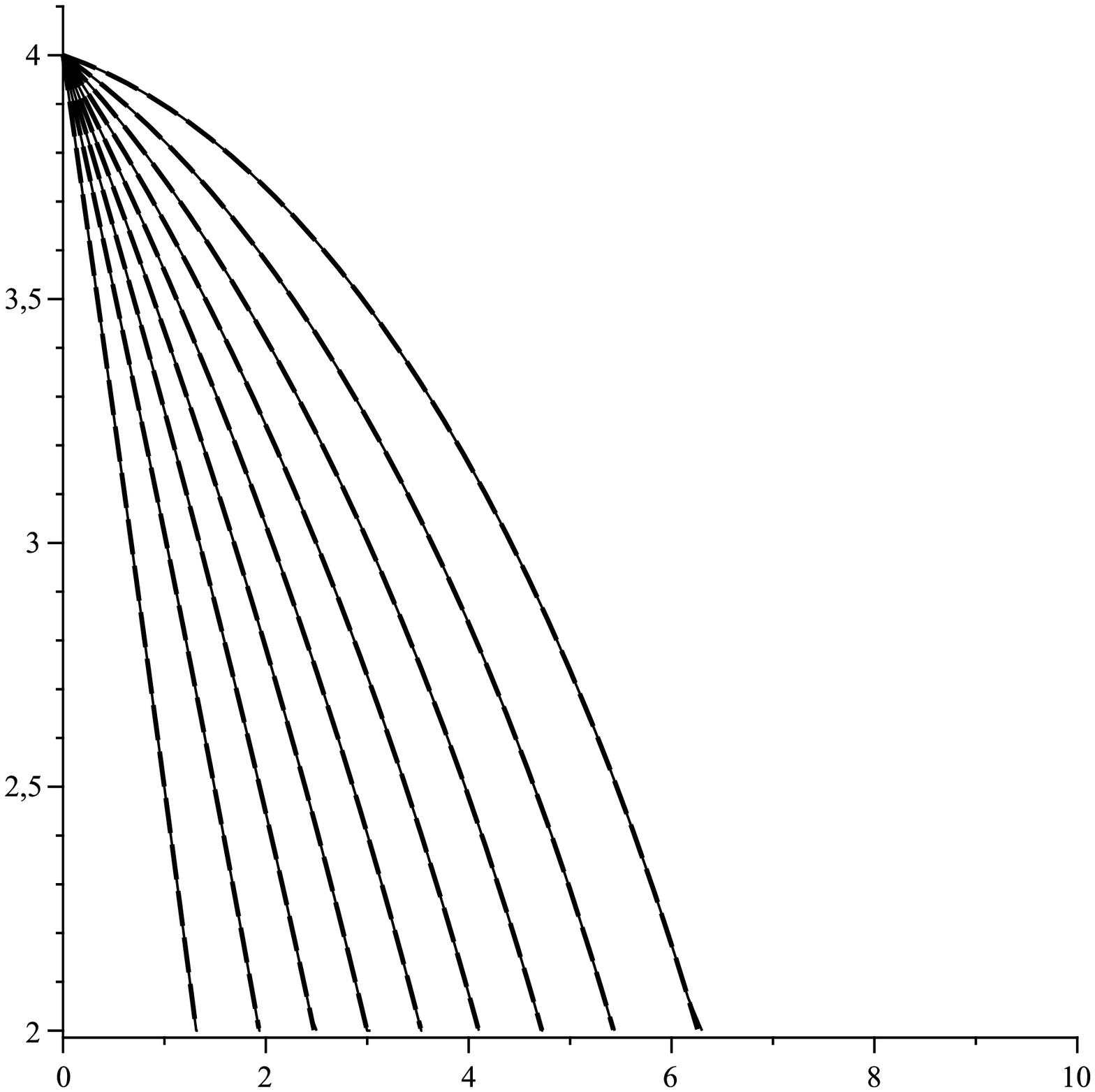}\\[.4cm]
\quad\mbox{(a)}\quad &\quad \mbox{(b)}
\end{array}$\\
\end{center}
\caption{The behavior of $r(\tau)$ is shown in the case of outgoing radiation with the same parameter choice as in Fig.~\ref{fig:4}, with axes given in units of $M_1$.
The initial conditions are $u(0)=-1000$, $r(0)=4$, $\phi(0)=0$, $\nu^{\hat \phi}(0)=0$ and purely radial motion with (a) $\nu^{\hat r}(0)=[0.1,0.2,0.3,0.4,0.5,0.6,0.7,0.8,0.9]$ (outward), (b) $\nu^{\hat r}(0)\to-\nu^{\hat r}(0)$ (inward).
In case (a) the bigger is the initial value of the velocity the longer is the interval of proper time spent by the particle before reaching the apparent horizon or eventually escaping outwards if the speed is large enough.
In case (b) it is just the opposite, with the particle always reaching the apparent horizon (faster and faster for $\nu^{\hat r}(0)$ increasingly negative).
}
\label{fig:8}
\end{figure}

In order to numerically integrate the equations of motion (\ref{motion2}) we must specify the mass function $M(u)$. 
We adopt the mass profile (\ref{massprofile}), so that the mass smoothly decreases/increases between the value $M_1$ corresponding to the past asymptotic Schwarzschild spacetime and the value $M_2$ of the future asymptotic Schwarzschild spacetime. 
Another possibility that we will not consider for numerical study would be a mass function which decreases/increases linearly with $u$ over a finite interval of time (see previous section). 

As typical solutions we consider the numerical examples shown in Figs.~\ref{fig:4}--\ref{fig:8}, where the geodesic behavior is compared with the motion of a particle interacting with the background radiation field in the case of outgoing radiation.
The same analysis can be easily repeated for the ingoing case.
We choose $\beta=10^{-2}$ and fix the value of the friction parameter obtained by dividing $\tilde\sigma$ by the mass $M_1$ of the past asymptotic Schwarzschild spacetime to be $10^4$, indicating a strong interaction of the test particle with the background radiation field. For small values of $\tilde\sigma$ deviations from geodesic motion are not significant.  

We have investigated two different conditions for the initial radius of the orbit: $r(0)=4M_1$ and $r(0)=6M_1$. 
The typical feature is that an initially circular orbit in the past asymptotic Schwarzschild spacetime spirals inwards if its velocity is smaller than the Keplerian one; instead for greater values  both the geodesic and accelerated particles escape outwards.
If the initial velocity equals the Keplerian one, instead, the geodesic particle escapes, whereas the accelerated one spirals towards the apparent horizon (see Fig.~\ref{fig:4}).
Increasing the initial value of the radius enriches the situation, as shown in Fig.~\ref{fig:6}.
In fact, the initially circular orbit in the past asymptotic Schwarzschild spacetime undergoes a transition to a quasi-circular geodesic in the future asymptotic Schwarzschild spacetime if its velocity equals the Keplerian one, i.e. the motion turns out to be confined in a region close to such a geodesic orbit, because the path oscillates between a minimum and a maximum radius.
If the interaction is not so strong the contribution of mass variation dominates with respect to that due to the acceleration, leading to an oscillating behavior of the accelerated orbit around an asymptotic radius as in in the case of geodesic motion (see Fig.~\ref{fig:7}).
Finally, if the path is initially radial, it remains radial, and the particle can eventually escape if it is directed outwards with a large enough initial speed (see Fig.~\ref{fig:8}).
Note that in this case geodesic and accelerated orbits are practically indistinguishable even for strong interaction.

It is worth to recall that in the original works on Poynting-Robertson effect the main concern was a situation outside normal star, where it is only relevant to consider the outgoing-radiation case. In the present paper, we are also
---actually mainly---interested in ultracompact centre like black hole, when the ingoing-radiation case 
is also relevant. Namely, a black hole does not itself radiate (if not taking quantum effects into account), 
it also does not allow any stable accretion configuration (as a radiating source) near the horizon, and 
finally, even if such a source was there, most of its radiation would fall below the horizon. Therefore, at 
least in the vicinity of the horizon the ingoing radiation occurs more probably than outgoing.

In the case of ingoing radiation the dominant effect for both geodesic and accelerated particles is a push towards the apparent horizon.
An initially circular orbit in the past asymptotic Schwarzschild spacetime always spirals inwards, eventually reaching the apparent horizon after a few revolutions in a finite proper time interval.  
The coupling with the background radiation field causes accelerated particles to cross the apparent horizon before the corresponding geodesics.
The radial motion is characterized by the same feature as in the case of outgoing radiation.

\section{Concluding remarks}

The Vaidya spacetime with a Thomson interaction of its null dust with test particle motion provides an arena for the investigation of a Poynting-Robertson-like effect in a self-consistent way without the requirement that the null dust itself be a test field. 
One should admit that using the exact, Vaidya solution, where the radiation flux is tied consistently 
to the mass loss/gain by the centre, has only a theoretical importance in most situations, because the 
mass change is almost always negligible with respect to the mass itself. (This may only be false in final 
stages of black hole evaporation.) However, the properties of this effect evident in the simpler case of a test radiation field in the Schwarzschild spacetime are reflected by those of the appropriate limit of the Vaidya case, but the latter case allows one to see how they change under more extreme conditions where the outgoing radiation itself contributes to the gravitational field.

\section*{Acknowledgement}

All authors thank ICRANet for support.
OS also thanks for support from Czech projects GACR-202/09/0772, MSM0021610860 and LC06014.

\end{document}